\documentclass[letterpaper,11pt]{article}
\usepackage{amsmath,mathrsfs,amssymb,amsfonts,amsthm,mathtools}
\usepackage{physics,color}
\usepackage{graphicx}
\usepackage{subfigure}
\usepackage{tabularx}
\usepackage[table]{xcolor}
\usepackage{tikz,tikz-cd}
\usetikzlibrary{shapes.geometric, arrows, positioning}
\usepackage{empheq}
\usepackage{bm}

\pdfoutput=1
\usepackage{jheppub}
\usepackage[T1]{fontenc}

\newcommand{\nn}{\nonumber}

\def\ft#1#2{{\textstyle{\frac{\scriptstyle #1}{\scriptstyle #2} } }}

\hypersetup{
    colorlinks=true,
    linkcolor=blue,
    citecolor=blue,
    urlcolor=blue
}

\title{Phase-Induced Particle Creation in the Kappa-Gamma Vacuum}

\author[a]{Arash Azizi}

\affiliation[a]{{\it The Institute for Quantum Science and Engineering,
Texas A\&M University,\nn\\ College Station, TX 77843, U.S.A.}}

\emailAdd{sazizi@tamu.edu}

\abstract{We develop a two-parameter family of flat-spacetime modes labeled by a deformation scale $\kappa$ and a phase angle $\gamma$, extending the $\kappa$-plane wave framework to include complex squeezing. The resulting $\kappa\gamma$ basis provides a globally well-defined mode decomposition whose associated vacuum $|0_{\kappa,\gamma}\rangle$ is a continuous-mode phase-squeezed state: $\kappa$ fixes the squeezing magnitude, while $\gamma$ sets the squeezing angle in phase space. We identify \textit{phase-induced particle creation}, in which a relative phase mismatch $\Delta\gamma$ between observers generates a nontrivial particle spectrum governed by $(\kappa,\Delta\gamma)$ even when $\kappa$ is held fixed. We then derive the two reciprocal Bogoliubov maps: the $\kappa\gamma$ plane-wave operators in terms of $\kappa'$-Rindler operators, and conversely the $\kappa'$-Rindler operators in terms of $\kappa\gamma$ plane-wave operators, providing a closed algebraic bridge between these bases.   Finally, by analyzing the Wightman function we show that $|0_{\kappa,\gamma}\rangle$ is globally regular, with no singularities beyond those of the standard Minkowski vacuum.}

\begin{document}
\maketitle
\flushbottom

\section{Introduction}

Quantum field theory (QFT) in curved spacetime
\cite{Unruh1976, Fulling1973, Hawking1975, DeWitt1975PhysicsRep, Davies1975, Wald1975, Fulling1989, Birrell_Davies1982, Wald1994, Mukhanov2007, Witten22QFT}
has profoundly reshaped our understanding of quantum phenomena in non-inertial frames and gravitational backgrounds. A central lesson is that the notion of vacuum is not unique but observer-dependent: the same quantum state can appear empty to one observer and populated to another. This is exemplified by the Unruh effect \cite{Unruh1976}, where a uniformly accelerated detector in the Minkowski vacuum responds as if immersed in a thermal bath at the Unruh temperature $T_U$, and by Hawking radiation from black holes \cite{Hawking1975}. These ideas have deep implications for relativistic quantum information \cite{Landulfo2009, Mann2012RQI, Su2014communication, Foo2020teleportation, Tjoa2022teleportation} and quantum optics \cite{Scully2003, Scully2006, Svidzinsky2021PRR, Svidzinsky2021PRL}.

Motivated by this observer-dependence, recent work has introduced one-parameter families of deformed vacua---``$\kappa$ vacua''---defined algebraically by a $\kappa$-dependent choice of mode functions. The Klein--Gordon inner product selects the positive-norm sector, canonical quantization associates it with annihilation operators, and the vacuum is the state annihilated by all of them. Thus changing the mode basis changes the particle notion and, in general, the vacuum.

In the Rindler-adapted construction of \cite{Azizi2022Kappashort, Azizi2023JHEP}---motivated by the structure of Unruh modes~\cite{Unruh1976, UnruhWald1984}---one considers one-parameter families of linear combinations of standard Rindler modes. The corresponding annihilators define a family of $\kappa$-dependent vacua; here we refer to them as \emph{$\kappa$-Rindler modes} to distinguish this Rindler-based family from the later $\kappa$-plane-wave construction. The family interpolates smoothly between the standard limiting vacua,
\begin{equation}
\kappa \to 0 \;\Rightarrow\; \text{Rindler vacuum}, 
\qquad
\kappa = 1 \;\Rightarrow\; \text{Minkowski vacuum}.
\end{equation}
For intermediate $\kappa$, when the field is prepared in a $\kappa$-Rindler vacuum, a uniformly accelerated Unruh--DeWitt detector \cite{Unruh1976, Einstein100, Colosi2009Rovelli} exhibits a thermal response with $\kappa$-dependent effective temperature
\begin{equation}
T_\kappa=\kappa T_U,
\end{equation}
as shown in \cite{Azizi2025Tunable}. This provides a controlled setting in which the spectral temperature is tunable by the choice of vacuum, while preserving the usual Rindler and Minkowski limits.

Complementary to this, the \emph{$\kappa$-plane-wave modes} were introduced in \cite{Azizi2025KappaPW}. One constructs a $\kappa$-dependent set of modes as linear combinations of standard Minkowski plane waves (hence the name), and then quantizes with respect to the Klein--Gordon positive-norm sector of this plane-wave-based basis. The resulting $\kappa$-vacuum is a continuous-mode squeezed state of the Minkowski vacuum, characterized by a \emph{real} squeezing parameter. In this setting, $\kappa$ controls the magnitude of the mode-by-mode squeezing, and the Minkowski vacuum is recovered smoothly in the limit $\kappa\to0$ within the same plane-wave organization. Together, the $\kappa$-Rindler and $\kappa$-plane-wave constructions provide two complementary perspectives on deformed vacua: one emphasizing tunable thermality, the other emphasizing real squeezing.

The natural next step, taken in the present work, is to allow the squeezing to be fully \emph{complex}. We introduce a second parameter, $\gamma$, into the $\kappa$-plane-wave construction, thereby defining a two-parameter family of \emph{$\kappa\gamma$ plane-wave modes} and vacua $\ket{0_{\kappa,\gamma}}$. In close analogy with single-mode squeezing in quantum optics, the deformation parameter $\kappa$ sets the squeezing magnitude, while the new parameter $\gamma$ fixes the squeezing \emph{angle} in phase space. The absolute phase of a single vacuum can be absorbed into a redefinition of the field modes, but the \emph{relative} phase difference $\Delta\gamma = \gamma' - \gamma$ between two vacua is physically meaningful and manifests itself in particle creation.

At this stage, the $\kappa$-deformations have been introduced in a purely algebraic way: one postulates a deformed mode basis and relates it to standard bases by Bogoliubov transformations, which is sufficient to define a consistent notion of particles and vacuum. Nevertheless, it is important to supply a concrete dynamical setting in which such vacua arise from a physical mechanism. A particularly illuminating context is provided by radiation from moving mirrors, where time-dependent boundary conditions generate quantum fluxes and squeezed Gaussian states at future null infinity. Foundational contributions include the variable-length cavity of Moore \cite{Moore1970}, the analyses of conformal anomalies and mirror radiation by Fulling and Davies \cite{Fulling_Davies1976, Fulling_Davies1977}, and the Carlitz--Willey trajectory producing an exactly thermal spectrum \cite{Carlitz_Willey1987}. Later developments clarified dynamical Casimir radiation and Hamiltonian formulations \cite{Ford1982mirror, Law1995}, quantified time-dependent particle creation \cite{Good2013mirror}, and surveyed broader dynamical Casimir and cosmological particle-production scenarios \cite{Dodonov2020review, Ford2021review}.

Within this framework, two companion manuscripts provide explicit \emph{boundary-driven} realizations of the algebraic vacua considered here. The $\kappa$ \emph{plane-wave} vacuum is realized by an accelerating mirror that reproduces the translation-invariant squeeze structure of the $\kappa$-vacuum on $\mathscr{I}^+$ \cite{Azizi2025Mirror}. The present $\kappa\gamma$ \emph{plane-wave} vacuum is realized by a \emph{modulated} accelerating mirror whose weak, frequency-diagonal, time-dependent Robin impedance rotates the squeezing angle while keeping the Planck weights fixed \cite{Azizi2025ModMirror}. In what follows, we therefore distinguish between the \emph{algebraic} construction of $\kappa\gamma$ vacua, carried out here, and their \emph{physical} realization via modulated mirrors.

With this physical picture in mind, the present paper provides a systematic algebraic characterization of the $\kappa\gamma$-vacuum and its observable consequences. Our main results are as follows. First, we identify a phenomenon we call \emph{phase-induced particle creation}: for fixed $\kappa$, a change in the phase parameter from $\gamma$ to $\gamma'$ induces a non-trivial Bogoliubov transformation between the corresponding operator bases, so that observers associated with different phases generally disagree on particle content. This effect is encoded in a particle number spectrum
\begin{equation}
N_\Lambda(\gamma',\gamma) = \frac{\sin^2(\gamma' - \gamma)}{\sinh^2(\pi\Lambda/\kappa)},
\end{equation}
which depends only on the relative phase $\Delta\gamma$ and vanishes when the phases coincide. Importantly, this does not alter the Unruh effect: when two observers are associated with the same $(\kappa,\gamma)$ vacuum, they agree on the particle content, and in the appropriate limit one recovers the standard Unruh temperature $T_U$ for the Minkowski vacuum.

Second, we show that the vacuum $\ket{0_{\kappa,\gamma}}$ is a continuous-mode \emph{phase-squeezed} Gaussian state of the Minkowski vacuum in the sense of quantum optics \cite{Scully_Zubairy1997} and continuous-variable quantum information \cite{Braunstein2005, Weedbrook2012}. For each frequency mode $\Lambda$, it is characterized by the condition
\begin{equation}
\left( a_\Lambda - e^{-\frac{\pi \Lambda}{\kappa}} e^{2i\gamma} a^\dagger_\Lambda \right) \ket{0_{\kappa,\gamma}} = 0,
\end{equation}
so that $\kappa$ controls the squeezing magnitude and $\gamma$ fixes the squeezing axis in phase space. This generalizes the purely real squeezing analysed in \cite{Azizi2025KappaPW}.

\begin{table}[h!]
\centering
\caption{Summary of the $\kappa$-mode families and their key properties.}
\arrayrulecolor{black}
\setlength{\arrayrulewidth}{0.3mm}
\renewcommand{\arraystretch}{1.8}
\begin{tabular}{|>{\columncolor{gray!15}}m{2.6cm}|
                 >{\centering\arraybackslash}m{1.7cm}|
                 >{\centering\arraybackslash}m{4cm}|
                 >{\centering\arraybackslash}m{5.5cm}|}
\hline
\rowcolor{blue!18}
\textbf{Mode Family} & \textbf{Operator} & \textbf{Special Limits} & \textbf{Key Physical Feature} \\
\hline
\textit{Kappa Rindler}
& ${\cal B}_{\Omega,\kappa}$
\newline $(\Omega \in \mathbb{R})$
& $\kappa \to 0$: Rindler vac.
\newline $\kappa = 1$: Minkowski vac.
& Tunable Thermal Response ($T_\kappa = \kappa T_U$)
\\
\hline
\textit{Kappa \newline Plane Wave}
& ${\cal A}_{\Lambda,\kappa}$
\newline $(\Lambda \in \mathbb{R}^+)$
& $\kappa \to 0$: Minkowski vac.
& Continuous real squeezing
\\
\hline
\textit{Kappa-Gamma \quad Plane Wave}
& ${\cal A}_{\Lambda,\kappa,\gamma}$
\newline $(\Lambda \in \mathbb{R}^+)$
& $\gamma = 0$: reduces to \textit{$\kappa$-plane wave}
\newline $\kappa \to 0$: Minkowski vac.
& Phase-induced particle creation
\newline Continuous phase squeezing
\\
\hline
\end{tabular}
\label{tab:kappa_modes_summary}
\end{table}

Third, we derive generalized Bogoliubov transformations that connect the $\kappa\gamma$ plane-wave basis to the $\kappa$-Rindler basis and, as limiting cases, to the standard Minkowski and Rindler modes. These ``mother'' transformations provide a unified framework for understanding how thermality, squeezing magnitude, and squeezing phase arise in different mode decompositions and observational settings. 

As summarized in Table~\ref{tab:kappa_modes_summary}, for the $\kappa$-Rindler family the deformation parameter does not modify the Unruh effect itself, but rather labels a set of additional vacua with tunable effective temperature. A uniformly accelerated detector in the Minkowski vacuum always responds thermally at the standard Unruh temperature $T_U$, independent of $\kappa$. By contrast, when the field is prepared in a $\kappa$-Rindler vacuum, the same detector sees an effective temperature $T_\kappa = \kappa T_U$, so that $\kappa = 1$ reproduces the usual Unruh temperature while the limit $\kappa \to 0$ corresponds to the Rindler vacuum with $T_\kappa \to 0$, i.e.\ an exactly empty state from the Rindler observer's point of view.

The Wightman function $W_{\kappa\gamma}(x,x')=\bra{0_{\kappa,\gamma}}\Phi(x)\Phi(x')\ket{0_{\kappa,\gamma}}$ is the central object controlling the local physics of the state: it determines detector response functions and, via point-splitting, fixes renormalized local observables such as $\langle T_{\mu\nu}\rangle$. In particular, the ultraviolet (Hadamard) behavior of the state is encoded entirely in the coincidence structure of $W$, so an explicit evaluation directly diagnoses whether the standard Minkowski subtraction suffices. For the $\kappa\gamma$ vacuum one has the usual chiral split $W_{\kappa\gamma}(x,x')=W^{\rm RTW}_{\kappa\gamma}(u,u')+W^{\rm LTW}_{\kappa\gamma}(v,v')$, where RTW (LTW) denote the right (left)-travelling wave sector, and \newline $W^{\rm RTW}_{\kappa\gamma}(u,u')=W^{\rm RTW}_{\rm th}(\Delta u)+W^{\rm RTW}_{\rm ph}(\Sigma;\gamma)$ with $\Delta u=u-u'$ and $\Sigma=u+u'$ (and analogously for LTW). The thermal KMS~\cite{Martin_Schwinger1959, Kubo1957} channel $W^{\rm RTW}_{\rm th}(\Delta u)$ reproduces the standard logarithmic coincidence singularity $-\frac{1}{4\pi}\ln\!\big[\mu^2(\Delta u-i\epsilon)\big]$ as $\Delta u\to0$, with $\kappa$ entering only through smooth even-power corrections beginning at $O((\Delta u)^2)$, while the phase-sensitive channel depends on the sum coordinate $\Sigma$ and is finite in the coincidence limit at fixed $\Sigma$: the even part is analytic through $\ln\cosh(\kappa\Sigma/2)$, and the odd part is finite after subtracting the universal IR divergence, with the residual scheme dependence encoded by the IR scale $\mu_{\rm IR}$. Consequently one may write $W_{\kappa\gamma}(x,x')=W_{\rm M}(x,x')+F_{\kappa\gamma}(x,x')$ with $F_{\kappa\gamma}$ smooth near $x=x'$, confirming that the $\kappa\gamma$ vacuum is Hadamard (its UV singularity is exactly Minkowskian) and that no additional divergences arise on the null lines $u=0$ or $v=0$ beyond the usual light-cone coincidence behavior.

The paper is organized as follows. Section~\ref{sec:kappa_gamma_modes} introduces the $\kappa\gamma$ plane-wave modes and their inner-product structure. Section~\ref{sec:bogoliubov} derives the generalized Bogoliubov transformations connecting $(\kappa,\gamma)$ bases to one another, to Minkowski modes, and to the $\kappa$-Rindler family. Section~\ref{sec:wightman} evaluates the Wightman function of the $\kappa\gamma$ vacuum and decomposes it into a stationary (thermal) channel and a phase-sensitive channel. Section~\ref{sec:discussion} interprets $\ket{0_{\kappa,\gamma}}$ as a continuous-mode phase-squeezed state and connects $\kappa$ and $\gamma$ to a modulated accelerating-mirror realization. We conclude in Section~\ref{sec:conclusion}. Appendix~\ref{app:notation} fixes notation and operator/frequency conventions. The detailed intermediate steps of the Bogoliubov analysis are given in Appendix~\ref{app:derivation_kg_in_kr} (deriving ${\cal A}_{\Lambda,\kappa,\gamma}$ in terms of ${\cal B}_{\Omega,\kappa'}$) and Appendix~\ref{app:derivation_kr_in_kg} (deriving ${\cal B}_{\Omega,\kappa'}$ in terms of ${\cal A}_{\Lambda,\kappa,\gamma}$). The Gamma-function and regulator identities used throughout the projections are collected in Appendix~\ref{app:gamma}. For the Wightman-function evaluation, we list the auxiliary integral identities in Appendix~\ref{app:GR-integrals}. The small-$\Sigma$ expansion of $I_{\sin}^{\mathrm{(ren)}}$ is derived in Appendix~\ref{app:small-expansion}, and the polar decomposition of the complex parameter $\eta$ is given in Appendix~\ref{app:eta_polar}.

\section{Kappa-Gamma Plane Wave Modes} \label{sec:kappa_gamma_modes}
Throughout this section we work with a massless scalar field in $(1+1)$-dimensional Minkowski spacetime and focus on the right-moving sector. It is convenient to use the light-cone coordinate $u=t-x$, in terms of which right-movers depend only on $u$. At the level of mode functions, the most general right-moving combination built from plane waves can be written (temporarily allowing the frequency label to range over $\mathbb{R}$) as
\begin{equation}
\Phi(u,\Omega) = \alpha(\Omega)\, e^{-i \Omega u} + \beta(\Omega)\, e^{i \Omega u}\,. \label{mode}
\end{equation}
The complex coefficients are constrained by the requirement that the resulting family forms an orthonormal basis with respect to the Klein--Gordon inner product. Making these constraints explicit also clarifies why it is ultimately preferable to label the modes by strictly positive frequencies.

\subsection{Inner Product Constraints and Frequency Restriction}
We adopt the standard Klein--Gordon inner product on a $t=\text{const}$ Cauchy slice,
\begin{equation}
\langle \phi_1, \phi_2 \rangle = i \int dx \,\big(\phi_1^* \partial_t \phi_2 - \phi_2 \partial_t \phi_1^*\big)\,,
\end{equation}
which, when evaluated on right-moving plane waves, yields
\begin{equation}
\left\langle e^{-i \Omega u}, e^{-i \Omega' u} \right\rangle = 4\pi \Omega\, \delta(\Omega-\Omega')\,.
\end{equation}
We require the family of modes to satisfy
\begin{equation}
\langle \Phi(u, \Omega), \Phi(u, \Omega') \rangle = \delta(\Omega - \Omega') \quad \text{and} \quad \langle \Phi(u, \Omega), \Phi^*(u, \Omega') \rangle = 0\,.
\end{equation}
Substituting \eqref{mode} and using the plane-wave overlaps gives
\begin{align}
\langle\Phi(u, \Omega), \Phi(u, \Omega')\rangle
&= 4 \pi \Omega\, \delta(\Omega-\Omega')\big(|\alpha(\Omega)|^2 - |\beta(\Omega)|^2\big) \notag \\
&\quad + 4 \pi \Omega\, \delta(\Omega+\Omega') \Big(\alpha^*(\Omega)\, \beta(-\Omega) - \beta^*(\Omega)\, \alpha(-\Omega)\Big). \label{inn.prod1}
\end{align}
Matching to $\delta(\Omega-\Omega')$ yields two conditions: a normalization within each frequency sector, and a consistency condition associated with the $\delta(\Omega+\Omega')$ support,
\begin{align}
|\alpha(\Omega)|^2 - |\beta(\Omega)|^2 &= \frac{1}{4 \pi \Omega}\,, \label{constraint1} \\
\alpha^*(\Omega) \beta(-\Omega) - \beta^*(\Omega) \alpha(-\Omega) &= 0\,. \label{constraint2}
\end{align}

The second orthogonality requirement, $\langle \Phi(u, \Omega), \Phi^*(u, \Omega') \rangle = 0$, reads
\begin{align}
\langle \Phi(u, \Omega), \Phi^*(u, \Omega') \rangle &=
\alpha^*(\Omega)\beta^*(\Omega') \left\langle e^{-i \Omega u}, e^{-i \Omega' u} \right\rangle
+ \beta^*(\Omega)\alpha^*(\Omega') \left\langle e^{i \Omega u}, e^{i \Omega' u} \right\rangle \notag \\
&\quad + \alpha^*(\Omega)\alpha^*(\Omega') \left\langle e^{-i \Omega u}, e^{i \Omega' u} \right\rangle
+ \beta^*(\Omega)\beta^*(\Omega') \left\langle e^{i \Omega u}, e^{-i \Omega' u} \right\rangle.
\end{align}
After substituting the plane-wave inner products, the $\delta(\Omega-\Omega')$ terms cancel when $\Omega=\Omega'$, and one finds
\begin{equation}
\langle \Phi(u, \Omega), \Phi^*(u, \Omega') \rangle
= 4\pi \Omega\, \delta(\Omega+\Omega') \Big[ \alpha^*(\Omega)\alpha^*(-\Omega) - \beta^*(\Omega)\beta^*(-\Omega) \Big]\,.
\end{equation}
Enforcing orthogonality therefore requires the bracket to vanish, which can be written equivalently as
\begin{equation}
-\alpha(\Omega)\alpha(-\Omega) + \beta(\Omega)\beta(-\Omega) = 0\,. \label{constraint3}
\end{equation}

The combined content of \eqref{constraint2} and \eqref{constraint3} is that if $\Omega$ is allowed to take both signs, then the would-be modes labeled by $+\Omega$ and $-\Omega$ are tied together by the $\delta(\Omega+\Omega')$ channel. Moreover, for a genuinely mixed-frequency mode with $\alpha(\Omega)\neq0$ and $\beta(\Omega)\neq0$, \eqref{constraint2} and \eqref{constraint3} together force $|\alpha(\Omega)|^2=|\beta(\Omega)|^2$, which is incompatible with the normalization \eqref{constraint1}. A standard and cleaner resolution is therefore to restrict the mode label to strictly positive frequencies from the outset.

Introducing $\Lambda>0$ as the frequency parameter, all terms supported on $\delta(\Lambda+\Lambda')$ are automatically absent, and the additional constraints \eqref{constraint2} and \eqref{constraint3} play no role. What remains is the single Bogoliubov-type normalization condition
\begin{equation}
|\alpha(\Lambda)|^2 - |\beta(\Lambda)|^2 = \frac{1}{4 \pi \Lambda}\,. \label{constraint_final}
\end{equation}
This condition is precisely what is needed for the associated ladder operators to satisfy canonical bosonic commutation relations.

\subsection{The Kappa-Gamma Ansatz and Field Expansion}
We now choose a concrete parametrization of the coefficients that satisfies \eqref{constraint_final} and, at the same time, introduces a controlled deformation away from ordinary Minkowski plane waves. The ansatz is expressed in terms of a real deformation scale $\kappa$ and a phase angle $\gamma$:
\begin{equation}
\alpha(\Lambda, \kappa, \gamma) = \frac{e^{\frac{\pi \Lambda}{2\kappa}} e^{i\gamma}}{\sqrt{{\cal N}_{\Lambda,\kappa}}} \,,
\qquad
\beta(\Lambda, \kappa, \gamma) = \frac{e^{-\frac{\pi \Lambda}{2\kappa}} e^{-i\gamma}}{\sqrt{{\cal N}_{\Lambda,\kappa}}}\,, \label{ansatz}
\end{equation}
with ${\cal N}_{\Lambda,\kappa} = 8\pi\Lambda\sinh\big(\frac{\pi\Lambda}{\kappa}\big)$. The parameter $\kappa$ controls the relative weight of the positive- and negative-frequency components, while $\gamma$ tracks their relative phase. The exponential asymmetry between the two terms is the hallmark of continuous-mode squeezing in the vacuum associated with this basis.

With \eqref{ansatz}, the corresponding normalized mode function becomes the $\kappa\gamma$ plane-wave mode
\begin{equation}
\Phi(u,\Lambda,\kappa,\gamma)= \frac{1}{{\sqrt{{\cal N}_{\Lambda,\kappa}}}}
\left( e^{\frac{\pi \Lambda}{2\kappa}} e^{i\gamma} e^{-i \Lambda u}
+ e^{-\frac{\pi \Lambda}{2\kappa}} e^{-i\gamma} e^{i \Lambda u} \right) \,.\label{kappagammamode}
\end{equation}
Given \eqref{constraint_final} and the definition \eqref{kappagammamode}, the associated annihilation and creation operators obey the standard bosonic algebra
\begin{equation}
[{\cal A}_{\Lambda,\kappa,\gamma},{\cal A}^\dagger_{\Lambda',\kappa,\gamma}]
=\delta(\Lambda-\Lambda'),\qquad
[{\cal A},{\cal A}]=[{\cal A}^\dagger,{\cal A}^\dagger]=0\,.
\end{equation}
Accordingly, the right-moving field admits the mode expansion
\begin{equation}
\Phi(u)=\int_{0}^{\infty} d\Lambda \Big[\Phi(u,\Lambda,\kappa,\gamma) {\cal A}_{\Lambda,\kappa,\gamma} + \Phi^*(u,\Lambda,\kappa,\gamma) {\cal A}^{\dagger}_{\Lambda,\kappa,\gamma} \Big]\,,
\end{equation}
and we define the corresponding vacuum by ${\cal A}_{\Lambda,\kappa,\gamma}|0_{\kappa,\gamma}\rangle = 0$ for all $\Lambda > 0$.

Ordinary Minkowski plane waves are recovered in an appropriate limit. For fixed $\Lambda>0$, taking $\kappa \to 0^+$ exponentially suppresses the negative-frequency component, leaving
\begin{equation}
\lim_{\kappa \to 0^+} \Phi(u, \Lambda, \kappa, \gamma) = \frac{e^{i\gamma}}{\sqrt{4 \pi \Lambda}} e^{-i \Lambda u} \,. \label{PlanewaveMink}
\end{equation}
Thus the $\kappa\gamma$ modes interpolate continuously to standard Minkowski right-movers, up to an overall phase. At the operator level, this corresponds to ${\cal A}_{\Lambda, 0, \gamma} = e^{-i\gamma} a_\Lambda$, with $a_\Lambda$ the usual Minkowski annihilation operator. The phase label $\gamma$ becomes physically meaningful when comparing descriptions that assign different values of $\gamma$ to the same underlying field configuration.

\subsection{Phase-Induced Particle Creation} \label{sec:phase_creation}
A consequence of introducing the phase parameter $\gamma$ is that changing $\gamma$ produces a nontrivial Bogoliubov map between operator bases. The deformation scale $\kappa$ is held fixed, but the relative phase between the positive- and negative-frequency components of the $\kappa\gamma$ mode changes, and this generically mixes annihilation and creation operators. To make this explicit, we compare two bases, $({\cal A}_{\Lambda,\kappa,\gamma}, {\cal A}^\dagger_{\Lambda,\kappa,\gamma})$ and $({\cal A}_{\Lambda,\kappa,\gamma'}, {\cal A}^\dagger_{\Lambda,\kappa,\gamma'})$, associated with the same $\kappa$ but with phases $\gamma$ and $\gamma'$.

The relation between these two bases is a special case of the generalized Bogoliubov transformations derived in Section~\ref{sec:bogoliubov}. Setting $\kappa'=\kappa$ in the general formulas yields
\begin{equation}
{\cal A}_{\Lambda, \kappa, \gamma'}
= \Big[ \cos(\gamma' - \gamma)
- i\coth\Big(\frac{\pi\Lambda}{\kappa}\Big)\sin(\gamma' - \gamma) \Big]
{\cal A}_{\Lambda, \kappa, \gamma}
- i \frac{\sin(\gamma' - \gamma)}{\sinh(\frac{\pi\Lambda}{\kappa})} {\cal A}^{\dagger}_{\Lambda, \kappa, \gamma}\,,
\label{eq:phase_bogoliubov_edited}
\end{equation}
which preserves the canonical commutation relations while explicitly mixing ${\cal A}$ and ${\cal A}^\dagger$. Consequently, the vacua $\ket{0_{\kappa,\gamma}}$ and $\ket{0_{\kappa,\gamma'}}$, defined by ${\cal A}_{\Lambda,\kappa,\gamma}\ket{0_{\kappa,\gamma}}=0$ and ${\cal A}_{\Lambda,\kappa,\gamma'}\ket{0_{\kappa,\gamma'}}=0$ for all $\Lambda>0$, correspond to inequivalent particle notions.

From the perspective of an observer employing the $(\kappa,\gamma)$ particle definition, the $(\kappa,\gamma')$ vacuum appears populated. The corresponding number density per mode is given by the modulus-squared of the Bogoliubov $\beta$ coefficient, namely the coefficient multiplying ${\cal A}^{\dagger}_{\Lambda, \kappa, \gamma}$ in Eq.~\eqref{eq:phase_bogoliubov_edited}:
\begin{equation}
N_\Lambda(\gamma', \gamma) = |\beta_{\Lambda\gamma\gamma'}|^2
= \frac{\sin^2(\gamma'-\gamma)}{\sinh^2(\frac{\pi\Lambda}{\kappa})}\,.
\label{eq:particle_number}
\end{equation}
This expression vanishes at $\Delta\gamma\equiv\gamma'-\gamma=0$ and is maximized at $\Delta\gamma=\pi/2$ (mod $\pi$). For fixed $\Lambda>0$, $N_\Lambda$ is an increasing function of $\kappa$, since $\sinh(\pi\Lambda/\kappa)$ decreases as $\kappa$ grows. For fixed $\kappa$, the distribution is largest in the infrared. In particular, the plane-wave labeling implies an infrared enhancement
\begin{equation}
N_\Lambda \sim \Big(\frac{\kappa^2}{\pi^2}\Big)\sin^2(\Delta\gamma)\,\Lambda^{-2}\,,
\qquad
\Lambda\to 0\,,
\end{equation}
so Eq.~\eqref{eq:particle_number} should be interpreted with an infrared regulator (for example, wavepackets or an explicit cutoff $\mu_{\rm IR}$) when discussing integrated particle number. Finally, in the limit $\kappa\to0^+$, $N_\Lambda$ is exponentially suppressed for any fixed $\Lambda>0$.

\section{Generalized Bogoliubov Transformations}
\label{sec:bogoliubov}

This section collects the Bogoliubov maps relating the operator bases used throughout the paper. Each map follows by equating the corresponding mode expansions of the field and then extracting operator relations by suitable Fourier (for plane waves) or Mellin (for $\kappa$-Rindler) projections; the intermediate steps are deferred to the appendices.

\subsection{Between Kappa-Gamma Plane Wave Modes}

The most general transformation relates a $(\kappa,\gamma)$ basis to a $(\kappa',\gamma')$ basis. One starts from two equivalent expansions of $\Phi(u)$,
\begin{align}
\Phi(u)
&= \int_0^\infty \frac{d\Lambda}{\sqrt{{\cal N}_{\Lambda,\kappa}}}\,
\Bigg\{
\Big( e^{\frac{\pi\Lambda}{2\kappa}+ i\gamma} e^{-i\Lambda u}
    +e^{-\frac{\pi\Lambda}{2\kappa}- i\gamma} e^{+i\Lambda u}\Big)
{\cal A}_{\Lambda,\kappa,\gamma}
\nn\\
&\hspace{3.1em}
+\Big( e^{\frac{\pi\Lambda}{2\kappa}- i\gamma} e^{+i\Lambda u}
     +e^{-\frac{\pi\Lambda}{2\kappa}+ i\gamma} e^{-i\Lambda u}\Big)
{\cal A}^{\dagger}_{\Lambda,\kappa,\gamma}
\Bigg\}
\nn\\
&= \int_0^\infty \frac{d\Lambda'}{\sqrt{{\cal N}_{\Lambda',\kappa'}}}\,
\Bigg\{
\Big( e^{\frac{\pi\Lambda'}{2\kappa'}+ i\gamma'} e^{-i\Lambda' u}
    +e^{-\frac{\pi\Lambda'}{2\kappa'}- i\gamma'} e^{+i\Lambda' u}\Big)
{\cal A}_{\Lambda',\kappa',\gamma'}
\nn\\
&\hspace{3.1em}
+\Big( e^{\frac{\pi\Lambda'}{2\kappa'}- i\gamma'} e^{+i\Lambda' u}
     +e^{-\frac{\pi\Lambda'}{2\kappa'}+ i\gamma'} e^{-i\Lambda' u}\Big)
{\cal A}^{\dagger}_{\Lambda',\kappa',\gamma'}
\Bigg\}.
\end{align}
Fourier projection with $\Lambda_0>0$ (using $\int_{-\infty}^{+\infty}du\,e^{i\Sigma u}=2\pi\delta(\Sigma)$) yields
\begin{align}
\int_{-\infty}^{+\infty}du\,e^{i\Lambda_0 u}\,\Phi(u)
&= \frac{2\pi}{\sqrt{{\cal N}_{\Lambda_0,\kappa}}}
\Big( e^{\frac{\pi\Lambda_0}{2\kappa}+ i\gamma}\,{\cal A}_{\Lambda_0,\kappa,\gamma}
    +e^{-\frac{\pi\Lambda_0}{2\kappa}+ i\gamma}\,{\cal A}^{\dagger}_{\Lambda_0,\kappa,\gamma}\Big)
\nn\\
&= \frac{2\pi}{\sqrt{{\cal N}_{\Lambda_0,\kappa'}}}
\Big( e^{\frac{\pi\Lambda_0}{2\kappa'}+ i\gamma'}\,{\cal A}_{\Lambda_0,\kappa',\gamma'}
    +e^{-\frac{\pi\Lambda_0}{2\kappa'}+ i\gamma'}\,{\cal A}^{\dagger}_{\Lambda_0,\kappa',\gamma'}\Big).
\end{align}
Taking the Hermitian conjugate of this relation and collecting both equations into a $2\times2$ system gives
\begin{align}
\begin{pmatrix}
e^{\frac{\pi\Lambda_0}{2\kappa}+i\gamma} & e^{-\frac{\pi\Lambda_0}{2\kappa}+i\gamma} \\
e^{-\frac{\pi\Lambda_0}{2\kappa}-i\gamma} & e^{\frac{\pi\Lambda_0}{2\kappa}-i\gamma}
\end{pmatrix}
\begin{pmatrix}
{\cal A}_{\Lambda_0,\kappa,\gamma}\\[2pt]
{\cal A}^{\dagger}_{\Lambda_0,\kappa,\gamma}
\end{pmatrix}
&=
\sqrt{\frac{\sinh(\frac{\pi\Lambda_0}{\kappa})}{\sinh(\frac{\pi\Lambda_0}{\kappa'})}}
\begin{pmatrix}
e^{\frac{\pi\Lambda_0}{2\kappa'}+i\gamma'} & e^{-\frac{\pi\Lambda_0}{2\kappa'}+i\gamma'} \\
e^{-\frac{\pi\Lambda_0}{2\kappa'}-i\gamma'} & e^{\frac{\pi\Lambda_0}{2\kappa'}-i\gamma'}
\end{pmatrix}
\begin{pmatrix}
{\cal A}_{\Lambda_0,\kappa',\gamma'}\\[2pt]
{\cal A}^{\dagger}_{\Lambda_0,\kappa',\gamma'}
\end{pmatrix}\!.
\end{align}
The determinant of the left matrix is $2\sinh(\frac{\pi\Lambda_0}{\kappa})$, and after inversion and simplification one arrives at the compact Bogoliubov form
\begin{align}
{\cal A}_{\Lambda,\kappa',\gamma'}
&=
\frac{\sinh\!\Big(\frac{\pi \Lambda}{2}\big(\frac{1}{\kappa'}+\frac{1}{\kappa}\big)-i(\gamma'-\gamma)\Big)}
{\sqrt{\sinh(\frac{\pi \Lambda}{\kappa})\,\sinh(\frac{\pi \Lambda}{\kappa'})}}
\,{\cal A}_{\Lambda,\kappa,\gamma}
\nn\\
&\hspace{1.2em}
+\frac{\sinh\!\Big(\frac{\pi \Lambda}{2}\big(\frac{1}{\kappa'}-\frac{1}{\kappa}\big)-i(\gamma'-\gamma)\Big)}
{\sqrt{\sinh(\frac{\pi \Lambda}{\kappa})\,\sinh(\frac{\pi \Lambda}{\kappa'})}}
\,{\cal A}^{\dagger}_{\Lambda,\kappa,\gamma}\,.
\label{eq:bogoliubov_kg_kg}
\end{align}
This expression generalizes the $\kappa$ plane-wave result of \cite{Azizi2025KappaPW} to the $\kappa\gamma$ family. Moreover, Eq.~\eqref{eq:phase_bogoliubov_edited} is recovered as the special case $\kappa'=\kappa$.

\subsection{Kappa-gamma plane wave and Minkowski modes}

A key special case is the transformation between a $\kappa\gamma$ mode and a standard Minkowski plane-wave mode. This is obtained by taking the limit $\kappa'\to0$ and setting $\gamma'=0$ in Eq.~\eqref{eq:bogoliubov_kg_kg}, identifying ${\cal A}_{\Lambda,0,0}=a_\Lambda$, and then inverting the resulting single-mode Bogoliubov relation. The result can be written as
\begin{equation}
{\cal A}_{\Lambda,\kappa,\gamma}
=
\frac{1}{\sqrt{2\sinh(\tfrac{\pi\Lambda}{\kappa})}}
\left(
e^{\frac{\pi\Lambda}{2\kappa}}e^{-i\gamma}\,a_\Lambda
-
e^{-\frac{\pi\Lambda}{2\kappa}}e^{+i\gamma}\,a^\dagger_\Lambda
\right).
\label{eq:kg_mink_transf}
\end{equation}
This relation underlies the squeezed nature of the $\ket{0_{\kappa,\gamma}}$ vacuum, analyzed in Sec.~\ref{sec:discussion}.

\subsection{The ``mother of all Bogoliubov transformations''}

The most general maps connecting the $\kappa\gamma$ plane-wave basis to the $\kappa$-Rindler basis of Ref.~\cite{Azizi2023JHEP} are summarized next. These relations unify, as special cases, all mode decompositions used in this work for a massless scalar field in flat spacetime. The derivations appear in Apps.~\ref{app:derivation_kg_in_kr} and \ref{app:derivation_kr_in_kg}.

First, the $\kappa\gamma$ plane-wave operator can be written in terms of $\kappa'$-Rindler operators as
\begin{align}
{\cal A}_{\Lambda,\kappa,\gamma}
&= -\frac{i}{2\pi}\int_{-\infty}^{\infty} d\Omega\,
\frac{1}{\sqrt{\Lambda\,\Omega\,\sinh(\frac{\pi\Lambda}{\kappa})\,\sinh(\frac{\pi\Omega}{\kappa'})}}
\nn\\
&\quad\times \Bigg\{
\Lambda^{-i\Omega}\Gamma(1+i\Omega)\,
\Bigg[
e^{\frac{\pi\Lambda}{2\kappa}-i\gamma}\,
\sinh\!\left(\frac{\pi\Omega}{2}\Big(\frac{1}{\kappa'}+1\Big)\right)
+
e^{-\frac{\pi\Lambda}{2\kappa}+i\gamma}\,
\sinh\!\left(\frac{\pi\Omega}{2}\Big(\frac{1}{\kappa'}-1\Big)\right)
\Bigg]
{\cal B}_{\Omega,\kappa'}
\nn\\
&\qquad\quad
+\Lambda^{+i\Omega}\Gamma(1-i\Omega)\,
\Bigg[
e^{\frac{\pi\Lambda}{2\kappa}-i\gamma}\,
\sinh\!\left(\frac{\pi\Omega}{2}\Big(\frac{1}{\kappa'}-1\Big)\right)
+
e^{-\frac{\pi\Lambda}{2\kappa}+i\gamma}\,
\sinh\!\left(\frac{\pi\Omega}{2}\Big(\frac{1}{\kappa'}+1\Big)\right)
\Bigg]
{\cal B}^{\dagger}_{\Omega,\kappa'}
\Bigg\}.
\label{KGPW-in-KR}
\end{align}
Conversely, the $\kappa'$-Rindler operator can be expressed in terms of $\kappa\gamma$ plane-wave operators as
\begin{align}
{\cal B}_{\Omega,\kappa'}
&=\frac{\operatorname{sgn}(\Omega)}{2\pi}\int_{0}^{\infty} d\Lambda\,
\sqrt{\frac{\Omega}{\sinh(\frac{\pi\Omega}{\kappa'})\,\Lambda\,\sinh(\frac{\pi\Lambda}{\kappa})}}\;
\Lambda^{i\Omega}\Gamma(-i\Omega)
\nn\\
&\quad\times \Bigg\{
\Bigg[
e^{\frac{\pi\Lambda}{2\kappa}+i\gamma}\,
\sinh\!\left(\frac{\pi\Omega}{2}\Big(\frac{1}{\kappa'}+1\Big)\right)
+
e^{-\frac{\pi\Lambda}{2\kappa}-i\gamma}\,
\sinh\!\left(\frac{\pi\Omega}{2}\Big(\frac{1}{\kappa'}-1\Big)\right)
\Bigg]{\cal A}_{\Lambda,\kappa,\gamma}
\nn\\
&\qquad\quad
+
\Bigg[
e^{\frac{\pi\Lambda}{2\kappa}-i\gamma}\,
\sinh\!\left(\frac{\pi\Omega}{2}\Big(\frac{1}{\kappa'}-1\Big)\right)
+
e^{-\frac{\pi\Lambda}{2\kappa}+i\gamma}\,
\sinh\!\left(\frac{\pi\Omega}{2}\Big(\frac{1}{\kappa'}+1\Big)\right)
\Bigg]{\cal A}^{\dagger}_{\Lambda,\kappa,\gamma}
\Bigg\}.
\label{KR-in-KGPW}
\end{align}
Together, Eqs.~\eqref{KGPW-in-KR} and \eqref{KR-in-KGPW} provide the master Bogoliubov map between the $\kappa\gamma$ plane-wave and $\kappa$-Rindler descriptions, from which the remaining special cases used in this work follow immediately.

\section{Detailed evaluation of the Wightman function}
\label{sec:wightman}
In this section we derive the right--moving Wightman function for the $\kappa\gamma$ state directly from the chiral mode expansion. The goal is to make every step explicit: the normalization of the modes, the distributional manipulations behind the mode sum, and the origin of the infrared (IR) subtleties that accompany a massless field in $(1+1)$ dimensions.
The right--moving field operator can be written as
\begin{align}
\Phi_{\rm RTW}(u)
&=
\int_0^\infty d\Lambda\,
\Big[
  \Phi_{\Lambda,\kappa,\gamma}(u)\,
  {\cal A}_{\Lambda,\kappa,\gamma}
 +\Phi^{\ast}_{\Lambda,\kappa,\gamma}(u)\,
  {\cal A}^{\dagger}_{\Lambda,\kappa,\gamma}
\Big],
\end{align}
where ${\cal A}_{\Lambda,\kappa,\gamma}$ and ${\cal A}^{\dagger}_{\Lambda,\kappa,\gamma}$ are the annihilation and creation operators associated with the $\kappa\gamma$ modes. They satisfy the canonical commutation relations
\begin{equation}
\big[{\cal A}_{\Lambda,\kappa,\gamma},
   {\cal A}^{\dagger}_{\Lambda',\kappa,\gamma}\big]
=\delta(\Lambda-\Lambda'),
\qquad
\big[{\cal A}_{\Lambda,\kappa,\gamma},
   {\cal A}_{\Lambda',\kappa,\gamma}\big]
=\big[{\cal A}^{\dagger}_{\Lambda,\kappa,\gamma},
   {\cal A}^{\dagger}_{\Lambda',\kappa,\gamma}\big]
=0,
\end{equation}
with all other commutators vanishing. The $\kappa\gamma$ vacuum is characterized by the condition
\begin{equation}
{\cal A}_{\Lambda,\kappa,\gamma}\,|0_{\kappa,\gamma}\rangle=0
\qquad
\text{for all }\Lambda>0.
\end{equation}
The corresponding right--moving Wightman function is defined by
\begin{equation}
W^{\rm RTW}_{\kappa\gamma}(u,u')
=\big\langle 0_{\kappa\gamma}\big|
\Phi_{\rm RTW}(u)\,\Phi_{\rm RTW}(u')
\big|0_{\kappa\gamma}\big\rangle.
\end{equation}
Substituting the mode expansion of $\Phi_{\rm RTW}$ yields
\begin{align}
W^{\rm RTW}_{\kappa\gamma}(u,u')
&=
\Big\langle 0_{\kappa\gamma}\Big|
\int_{0}^{\infty}\! d\Lambda\,
\Big[
  \Phi_{\Lambda,\kappa,\gamma}(u)\,
  {\cal A}_{\Lambda,\kappa,\gamma}
 +\Phi^{\ast}_{\Lambda,\kappa,\gamma}(u)\,
  {\cal A}^{\dagger}_{\Lambda,\kappa,\gamma}
\Big]\nn\\
&\hspace{5em}\times
\int_{0}^{\infty}\! d\Lambda'\,
\Big[
  \Phi_{\Lambda',\kappa,\gamma}(u')\,
  {\cal A}_{\Lambda',\kappa,\gamma}
 +\Phi^{\ast}_{\Lambda',\kappa,\gamma}(u')\,
  {\cal A}^{\dagger}_{\Lambda',\kappa,\gamma}
\Big]
\Big|0_{\kappa\gamma}\Big\rangle.
\end{align}
At this stage, most terms vanish immediately. Any contribution in which an annihilation operator acts on the ket vacuum is zero, and likewise any term in which a creation operator acts on the bra vacuum is zero. The only surviving contraction therefore pairs an annihilation operator with a creation operator, giving
\begin{align}
W^{\rm RTW}_{\kappa\gamma}(u,u')
&=
\int_{0}^{\infty}\! d\Lambda
\int_{0}^{\infty}\! d\Lambda'\,
\Phi_{\Lambda,\kappa,\gamma}(u)\,
\Phi^{\ast}_{\Lambda',\kappa,\gamma}(u')\,
\big\langle 0_{\kappa\gamma}\big|
{\cal A}_{\Lambda,\kappa,\gamma}
{\cal A}^{\dagger}_{\Lambda',\kappa,\gamma}
\big|0_{\kappa\gamma}\big\rangle.
\end{align}
Using the commutator to evaluate the vacuum expectation value,
\begin{equation}
\big\langle 0_{\kappa\gamma}\big|
{\cal A}_{\Lambda,\kappa,\gamma}
{\cal A}^{\dagger}_{\Lambda',\kappa,\gamma}
\big|0_{\kappa\gamma}\big\rangle
=
\big\langle 0_{\kappa\gamma}\big|
\big[{\cal A}_{\Lambda,\kappa,\gamma},
     {\cal A}^{\dagger}_{\Lambda',\kappa,\gamma}\big]
\big|0_{\kappa\gamma}\big\rangle
=
\delta(\Lambda-\Lambda'),
\end{equation}
the double integral collapses and we arrive at the standard mode--sum representation
\begin{align}
W^{\rm RTW}_{\kappa\gamma}(u,u')
&=
\int_{0}^{\infty}\! d\Lambda\,
\Phi_{\Lambda,\kappa,\gamma}(u)\,
\Phi^{\ast}_{\Lambda,\kappa,\gamma}(u').
\end{align}
This form relies only on the canonical commutation relations and the defining property of the vacuum; it does not depend on any special features of the $\kappa\gamma$ modes beyond completeness and orthonormality.
For the $\kappa\gamma$ state, a convenient normalized choice of right--moving mode functions is
\begin{equation}
\Phi_{\Lambda,\kappa,\gamma}(u)
=\frac{1}{\sqrt{{\cal N}_{\Lambda,\kappa}}}
\bigg(
  e^{\frac{\pi\Lambda}{2\kappa}+i\gamma}\,e^{-i\Lambda u}
 +e^{-\frac{\pi\Lambda}{2\kappa}-i\gamma}\,e^{+i\Lambda u}
\bigg),
\qquad
{\cal N}_{\Lambda,\kappa}
=8\pi\,\Lambda\,\sinh\!\Big(\frac{\pi\Lambda}{\kappa}\Big),
\end{equation}
where ${\cal N}_{\Lambda,\kappa}$ is fixed so that these modes are orthonormal with respect to the chiral Klein--Gordon inner product. Substituting into the mode sum gives
\begin{align}
W^{\rm RTW}_{\kappa\gamma}(u,u')
&=\int_{0}^{\infty}\! 
\frac{d\Lambda}{{\cal N}_{\Lambda,\kappa}} \bigg(
  e^{\frac{\pi\Lambda}{2\kappa}+i\gamma}\,e^{-i\Lambda u}
 +e^{-\frac{\pi\Lambda}{2\kappa}-i\gamma}\,e^{+i\Lambda u}
\bigg)
\bigg(
  e^{\frac{\pi\Lambda}{2\kappa}-i\gamma}\,e^{+i\Lambda u'}
 +e^{-\frac{\pi\Lambda}{2\kappa}+i\gamma}\,e^{-i\Lambda u'}
\bigg)\nn\\[1ex]
&= \frac{1}{8\pi}\int_{0}^{\infty}
\frac{d\Lambda}{\Lambda\,\sinh\!\big(\tfrac{\pi\Lambda}{\kappa}\big)}
\bigg(
   e^{\frac{\pi\Lambda}{\kappa}} e^{-i\Lambda(u-u')}
 + e^{-\frac{\pi\Lambda}{\kappa}} e^{i\Lambda(u-u')} \nn\\
 &\phantom{= \frac{1}{8\pi}\int_{0}^{\infty}
\frac{d\Lambda}{\Lambda\,\sinh\!\big(\tfrac{\pi\Lambda}{\kappa}\big)}
\bigg(}
 + e^{2i\gamma} e^{-i\Lambda(u+u')}
 + e^{-2i\gamma} e^{i\Lambda(u+u')}
\bigg).
\label{eq:W-rtw-start-app-rewritten}
\end{align}
This representation makes transparent the two distinct structures that enter the correlator: a stationary contribution that depends only on the null separation, and a phase--sensitive contribution that depends on the null sum variable.
We now introduce the shorthand
\begin{equation}
\Delta u=u-u',
\qquad
\Sigma=u+u',
\end{equation}
and treat separately the terms that depend only on $\Delta u$ (thermal, stationary channel) and those that depend on $\Sigma$ (phase--sensitive, non--stationary channel).

\subsection{\texorpdfstring{Thermal $\Delta u$-dependent contribution}{}}

We begin with the part of the correlator that depends only on the null separation $\Delta u=u-u'$. This is the stationary channel: it is invariant under a common shift \(u\to u+c\), \(u'\to u'+c\), and it reproduces the familiar KMS form. Concretely, we isolate the terms proportional to \(e^{\pm\pi\Lambda/\kappa}e^{\mp i\Lambda\Delta u}\) and introduce
\begin{equation}
I(\Delta u)
=\int_0^\infty
  \frac{d\Lambda}{\Lambda\sinh\!\big(\tfrac{\pi\Lambda}{\kappa}\big)}
  \Big[
     e^{\frac{\pi\Lambda}{\kappa}} e^{-i\Lambda\Delta u}
   + e^{-\frac{\pi\Lambda}{\kappa}} e^{i\Lambda\Delta u}
  \Big].
\label{eq:I-delta-def}
\end{equation}
The thermal contribution to the right--moving Wightman function is then
\begin{equation}
W^{\rm RTW}_{\rm th}(\Delta u)=\frac{1}{8\pi}\,I(\Delta u).
\end{equation}

To evaluate \(I(\Delta u)\), it is convenient to differentiate once with respect to \(\Delta u\), which removes the explicit \(1/\Lambda\) and leaves a standard oscillatory integral:
\begin{align}
\frac{\partial I}{\partial\Delta u}
&= -i\int_0^\infty
   \frac{d\Lambda}{\sinh\!\big(\tfrac{\pi\Lambda}{\kappa}\big)}
   \Big[
      e^{\frac{\pi\Lambda}{\kappa}} e^{-i\Lambda\Delta u}
    - e^{-\frac{\pi\Lambda}{\kappa}} e^{i\Lambda\Delta u}
   \Big].
\label{eq:dI-du-start}
\end{align}
Using
\begin{equation}
e^{\pm\frac{\pi\Lambda}{\kappa}}
=\cosh\!\Big(\frac{\pi\Lambda}{\kappa}\Big)
 \pm\sinh\!\Big(\frac{\pi\Lambda}{\kappa}\Big),
\end{equation}
we separate the integrand into a smooth thermal piece and a purely distributional contact term. A short algebraic rearrangement gives
\begin{align}
\frac{\partial I}{\partial\Delta u}
&= -2\int_0^\infty
      d\Lambda\,\coth\!\Big(\frac{\pi\Lambda}{\kappa}\Big)\sin(\Lambda\Delta u)
   - i\left(
        \int_0^\infty d\Lambda\,e^{i\Lambda\Delta u}
      + \int_0^\infty d\Lambda\,e^{-i\Lambda\Delta u}
      \right).
\label{eq:dI-du-decomp}
\end{align}

The first integral is a standard Gradshteyn--Ryzhik type result,
\begin{equation}
J_2(a,b)=\int_0^\infty d\Lambda\,\coth(a\Lambda)\,\sin(b\Lambda)
=\frac{\pi}{2a}\,\coth\!\Big(\frac{\pi b}{2a}\Big),
\qquad a>0,
\end{equation}
whose derivation we record in (\ref{app:J_2}). Taking \(a=\pi/\kappa\) and \(b=\Delta u\) yields
\begin{align}
\int_0^\infty d\Lambda\,\coth\!\Big(\frac{\pi\Lambda}{\kappa}\Big)\sin(\Lambda\Delta u)
&=\frac{\kappa}{2}\,
   \coth\!\Big(\frac{\kappa\Delta u}{2}\Big),
\end{align}
so that the thermal contribution from the first term in \eqref{eq:dI-du-decomp} becomes
\begin{equation}
-2\int_0^\infty
      d\Lambda\,\coth\!\Big(\frac{\pi\Lambda}{\kappa}\Big)\sin(\Lambda\Delta u)
=-\kappa\,\coth\!\Big(\frac{\kappa\Delta u}{2}\Big).
\end{equation}

The remaining term in \eqref{eq:dI-du-decomp} is supported only at coincidence. To display its distributional meaning, we introduce an exponential regulator,
\begin{equation}
\int_0^\infty d\Lambda\,e^{\pm i\Lambda\Delta u}
\;\longrightarrow\;
\int_0^\infty d\Lambda\,e^{\pm i\Lambda\Delta u}\,e^{-\epsilon\Lambda},
\qquad\epsilon>0,
\end{equation}
which gives
\begin{align}
\int_0^\infty d\Lambda\,e^{i\Lambda\Delta u}\,e^{-\epsilon\Lambda}
&=\frac{1}{\epsilon-i\Delta u},\\
\int_0^\infty d\Lambda\,e^{-i\Lambda\Delta u}\,e^{-\epsilon\Lambda}
&=\frac{1}{\epsilon+i\Delta u}.
\end{align}
Hence
\begin{align}
-i\left(
        \int_0^\infty d\Lambda\,e^{i\Lambda\Delta u}\,e^{-\epsilon\Lambda}
      + \int_0^\infty d\Lambda\,e^{-i\Lambda\Delta u}\,e^{-\epsilon\Lambda}
      \right)
&=-i\left(\frac{1}{\epsilon-i\Delta u}+\frac{1}{\epsilon+i\Delta u}\right) \nonumber\\
&=-i\,\frac{2\epsilon}{\epsilon^2+\Delta u^2}.
\end{align}
In the limit \(\epsilon\to 0^+\) this is the standard Lorentzian representation of a delta distribution,
\begin{equation}
\delta(\Delta u)
=\lim_{\epsilon\to0^+}\frac{1}{\pi}\,\frac{\epsilon}{\Delta u^2+\epsilon^2},
\end{equation}
so that, in the sense of distributions,
\begin{align}
-i\left(
        \int_0^\infty d\Lambda\,e^{i\Lambda\Delta u}
      + \int_0^\infty d\Lambda\,e^{-i\Lambda\Delta u}
      \right)
\;\longrightarrow\;-2\pi i\,\delta(\Delta u).
\end{align}
This contact term affects only the coincidence limit and is fixed by the same $i\epsilon$ prescription that defines the Wightman function.

Putting the two pieces together, we arrive at
\begin{equation}
\frac{\partial I}{\partial\Delta u}
=-\kappa\,\coth\!\Big(\frac{\kappa\Delta u}{2}\Big)
 -2\pi i\,\delta(\Delta u).
\label{eq:dI-du-final}
\end{equation}
Integrating with respect to \(\Delta u\) gives
\begin{equation}
\int d\Delta u\,\coth\!\Big(\frac{\kappa\Delta u}{2}\Big)
=\frac{2}{\kappa}\,\ln\Bigl|\sinh\!\Big(\frac{\kappa\Delta u}{2}\Big)\Bigr|,
\end{equation}
so, up to an integration constant,
\begin{equation}
I(\Delta u)
=-2\,\ln\Bigl|\sinh\!\Big(\frac{\kappa\Delta u}{2}\Big)\Bigr|
 -2\pi i\,\theta(\Delta u)+C.
\label{eq:I-Deltau-abs}
\end{equation}
The absolute value (and the step function) are fixed by the same \(i\epsilon\) prescription that defines the Wightman function. For small \(\epsilon>0\) we use the branch choice
\begin{equation}
\ln(x - i\epsilon)
 = \ln|x| - i\pi\bigl(1-\theta(x)\bigr).
\end{equation}
Accordingly,
\begin{equation}
\ln\!\Bigl[\sinh\!\Bigl(\frac{\kappa}{2}(\Delta u-i\epsilon)\Bigr)\Bigr]
=\ln\Bigl|\sinh\!\Bigl(\frac{\kappa\Delta u}{2}\Bigr)\Bigr|
 -i\pi\bigl[1-\theta(\Delta u)\bigr]
 +\mathcal{O}(\epsilon),
\end{equation}
and the discontinuous imaginary part from the logarithm accounts for the explicit \(-2\pi i\,\theta(\Delta u)\) term, leaving only a $\Delta u$--independent constant. Absorbing this constant into a renormalization scale \(\mu\), we may write
\begin{equation}
I(\Delta u)
=-2\,\ln\!\Big[\mu^{2}\,
\sinh\!\Big(\frac{\kappa}{2}(\Delta u-i\epsilon)\Big)\Big].
\end{equation}
The thermal part of the right--moving Wightman function is therefore
\begin{equation}
W^{\rm RTW}_{\rm th}(\Delta u)
=\frac{1}{8\pi}I(\Delta u)
=
-\frac{1}{4\pi}\,
\ln\!\Big[\mu^2\,\frac{2}{\kappa}\,
\sinh\!\Big(\frac{\kappa}{2}\,(\Delta u-i\epsilon)\Big)\Big],
\label{eq:Wth-RTW-app}
\end{equation}
which is the standard KMS thermal correlator at temperature \(T=\kappa/(2\pi)\).

\subsection{\texorpdfstring{Phase-sensitive $\Sigma$-dependence and IR behavior}{}}
\label{subsec:phase-Sigma-IR}

We now analyze the phase-sensitive terms, i.e.\ those proportional to $e^{\pm 2i\gamma}e^{\mp i\Lambda\Sigma}$ in \eqref{eq:W-rtw-start-app-rewritten}. It is natural to isolate the dependence on the sum coordinate $\Sigma=u+u'$ and to separate the result into parts that are even and odd under $\Sigma\to-\Sigma$.
Define
\begin{equation}
I'(\Sigma;\gamma)
\equiv
\int_{0}^{\infty}\frac{d\Lambda}{\Lambda\sinh(\pi\Lambda/\kappa)}
\Big(
  e^{2i\gamma}e^{-i\Lambda\Sigma}
 +e^{-2i\gamma}e^{i\Lambda\Sigma}
\Big),
\label{eq:Iprime-def}
\end{equation}
so that
\begin{equation}
W^{\rm RTW}_{\rm ph}(\Sigma;\gamma)
=\frac{1}{8\pi}\,I'(\Sigma;\gamma).
\end{equation}
Using
\[
e^{2i\gamma}e^{-i\Lambda\Sigma}
 +e^{-2i\gamma}e^{i\Lambda\Sigma}
=2\big[\cos(2\gamma)\cos(\Lambda\Sigma)+\sin(2\gamma)\sin(\Lambda\Sigma)\big],
\]
we split $I'$ as
\begin{align}
I'(\Sigma;\gamma)
&=2\cos(2\gamma)\,I_{\rm even}(\Sigma)
 +2\sin(2\gamma)\,I_{\rm odd}(\Sigma),
\label{eq:Iprime-even-odd-split}
\end{align}
with
\begin{align}
I_{\rm even}(\Sigma)
&=\int_0^\infty d\Lambda\,
   \frac{\cos(\Lambda\Sigma)}{\Lambda\sinh(\pi\Lambda/\kappa)},
\label{eq:I-even-def}\\
I_{\rm odd}(\Sigma)
&=\int_0^\infty d\Lambda\,
   \frac{\sin(\Lambda\Sigma)}{\Lambda\sinh(\pi\Lambda/\kappa)}.
\label{eq:I-odd-def}
\end{align}
By construction, $I_{\rm even}(-\Sigma)=I_{\rm even}(\Sigma)$ and $I_{\rm odd}(-\Sigma)=-I_{\rm odd}(\Sigma)$.

\subsubsection*{Even part: fixing the additive constant}

As written in \eqref{eq:I-even-def}, the even integral has an additive IR ambiguity: near $\Lambda=0$ one has $\sinh(\pi\Lambda/\kappa)\sim \pi\Lambda/\kappa$ and hence the integrand behaves as $\sim (\kappa/\pi)\Lambda^{-2}$, so the $\Sigma$-independent piece is ill-defined. Physically, this reflects the freedom to shift the Wightman function by a state-independent constant. We fix this freedom by imposing the renormalization condition $I_{\rm even}(0)=0$ (equivalently, by defining $I_{\rm even}$ with the replacement $\cos(\Lambda\Sigma)\to \cos(\Lambda\Sigma)-1$). With this understanding, it is convenient to differentiate once with respect to $\Sigma$, which removes the explicit $1/\Lambda$ and yields a convergent integral:
\begin{equation}
\frac{d I_{\rm even}}{d\Sigma}
=\int_0^\infty d\Lambda\,
\frac{-\sin(\Lambda\Sigma)}{\sinh(\pi\Lambda/\kappa)}.
\end{equation}
This is a standard Gradshteyn--Ryzhik \cite{Gradshteyn_Ryzhik2014} integral, whose derivation is recorded in (\ref{app:J_1}),
\begin{equation}
J_{1}(a,b)=\int_0^\infty d\Lambda\,\frac{\sin(b\Lambda)}{\sinh(a\Lambda)}
=\frac{\pi}{2a}\,\tanh\!\Big(\frac{\pi b}{2a}\Big),
\qquad a>0.
\end{equation}
Taking $a=\pi/\kappa$ and $b=\Sigma$ gives
\begin{equation}
\frac{d I_{\rm even}}{d\Sigma}
=-\frac{\kappa}{2}\,\tanh\!\Big(\frac{\kappa\Sigma}{2}\Big).
\end{equation}
Integrating in $\Sigma$ yields
\begin{align}
I_{\rm even}(\Sigma)
&=-\frac{\kappa}{2}\int d\Sigma\,\tanh\!\Big(\frac{\kappa\Sigma}{2}\Big)\\
&=-\ln\cosh\!\Big(\frac{\kappa\Sigma}{2}\Big)+C_{\rm even}.
\end{align}
Imposing $I_{\rm even}(0)=0$ fixes $C_{\rm even}=0$. The even contribution to $I'$ is therefore
\begin{equation}
I'_{\rm even}(\Sigma;\gamma)
=2\cos(2\gamma)\,I_{\rm even}(\Sigma)
=-2\cos(2\gamma)\,\ln\cosh\!\Big(\frac{\kappa\Sigma}{2}\Big),
\end{equation}
and the corresponding even phase-sensitive part of the Wightman function is
\begin{equation}
W^{\rm RTW}_{\rm ph,even}(\Sigma;\gamma)
=\frac{1}{8\pi}I'_{\rm even}(\Sigma;\gamma)
=-\frac{\cos(2\gamma)}{4\pi}\,
\ln\!\cosh\!\Big(\frac{\kappa\Sigma}{2}\Big).
\label{eq:W-phase-even-app-rew}
\end{equation}
This contribution is manifestly even in $\Sigma$ and finite for all finite $\Sigma$.

\subsubsection*{Odd part: IR divergence and renormalized sine sum}

We now turn to the odd part $I_{\rm odd}(\Sigma)$ in \eqref{eq:I-odd-def}. At the integrand level, the behavior near $\Lambda=0$ is already IR-singular:
\begin{equation}
\sinh\!\Big(\frac{\pi\Lambda}{\kappa}\Big)\sim\frac{\pi\Lambda}{\kappa},
\qquad
\sin(\Lambda\Sigma)\sim\Lambda\Sigma,
\end{equation}
so that
\begin{equation}
\frac{\sin(\Lambda\Sigma)}{\Lambda\sinh(\pi\Lambda/\kappa)}
\sim \frac{\kappa\Sigma}{\pi\,\Lambda},
\qquad \Lambda\to0,
\end{equation}
and therefore
\begin{equation}
I_{\rm odd}(\Sigma)
\sim \frac{\kappa\Sigma}{\pi}\int_0\frac{d\Lambda}{\Lambda},
\end{equation}
which exhibits a logarithmic IR divergence at the lower limit.

To reorganize this divergence it is convenient to use the standard series representation
\begin{equation}
\frac{1}{\sinh x}
=2\sum_{n=0}^\infty e^{-(2n+1)x}.
\end{equation}
Inserting this into \eqref{eq:I-odd-def} gives
\begin{align}
I_{\rm odd}(\Sigma)
&=2\sum_{n=0}^\infty\int_0^\infty d\Lambda\,
\frac{\sin(\Lambda\Sigma)}{\Lambda}\,
e^{-(2n+1)\pi\Lambda/\kappa}.
\end{align}
The inner integral is elementary:
\begin{equation}
\int_0^\infty d\Lambda\,
\frac{\sin(\Lambda\Sigma)}{\Lambda}\,e^{-a\Lambda}
=\arctan\!\Big(\frac{\Sigma}{a}\Big),
\qquad a>0,
\end{equation}
so with $a_n=(2n+1)\pi/\kappa$ we obtain
\begin{equation}
I_{\rm odd}(\Sigma)
=2\sum_{n=0}^\infty
\arctan\!\Big(\frac{\kappa\,\Sigma}{\pi(2n+1)}\Big).
\label{eq:Isin-series-naive-rew}
\end{equation}
For large $n$ one has $\arctan x \sim x$, so the tail behaves as
\begin{equation}
\arctan\!\Big(\frac{\kappa\,\Sigma}{\pi(2n+1)}\Big)
\sim \frac{\kappa\,\Sigma}{\pi(2n+1)},
\qquad n\gg1,
\end{equation}
and hence
\begin{equation}
I_{\rm odd}(\Sigma)
\sim\frac{2\kappa\Sigma}{\pi}\sum_{n=0}^\infty\frac{1}{2n+1},
\end{equation}
where the harmonic series over odd integers diverges logarithmically. This is the same IR divergence identified at the integral level, and it enters as a universal linear term in $\Sigma$.

Physically, different IR regulators correspond to different ways of cutting off the harmonic series; all such choices differ only by a contribution proportional to $\Sigma$. It is therefore natural to subtract this universal linear divergence and parameterize the remaining scheme dependence by an IR scale $\mu_{\rm IR}$.

To this end we define an IR-renormalized odd function
\begin{equation}
I_{\sin}^{\rm (ren)}(\Sigma;\mu_{\rm IR})
=
\frac{1}{2\pi}\sum_{n=0}^{\infty}
\!\left[
\arctan\!\frac{\kappa\,\Sigma}{\pi(2n{+}1)}
-
\frac{\kappa\,\Sigma}{\pi(2n{+}1)}
\right]
+\frac{\kappa\,\Sigma}{4\pi^2}\,
\ln\!\frac{\kappa}{2\pi\,\mu_{\rm IR}},
\label{eq:Isin-ren-def-rew}
\end{equation}
where:
\begin{itemize}
\item the subtraction in square brackets removes the large-$n$ tail $\propto\kappa\Sigma/(2n+1)$ and makes the sum absolutely convergent for all $\Sigma$;
\item the last term reinstates the scheme-dependent linear piece in $\Sigma$, encoded by the IR scale $\mu_{\rm IR}$.
\end{itemize}
By construction, $I_{\sin}^{\rm (ren)}(\Sigma;\mu_{\rm IR})$ is odd in $\Sigma$ and finite for all $\Sigma$.

From \eqref{eq:Iprime-even-odd-split} the odd contribution to $I'$ is
\begin{equation}
I'_{\rm odd}(\Sigma;\gamma)
=2\sin(2\gamma)\,I_{\rm odd}(\Sigma),
\end{equation}
and replacing $I_{\rm odd}$ by its renormalized version (up to the absorbed linear counterterm) is equivalent to writing
\begin{equation}
I'_{\rm odd}(\Sigma;\gamma)
=8\pi\,\sin(2\gamma)\,
I_{\sin}^{\rm (ren)}(\Sigma;\mu_{\rm IR}).
\end{equation}
Thus the odd phase-sensitive contribution to the Wightman function is
\begin{equation}
W^{\rm RTW}_{\rm ph,odd}(\Sigma;\gamma)
=\frac{1}{8\pi}I'_{\rm odd}(\Sigma;\gamma)
=\sin(2\gamma)\,I_{\sin}^{\rm (ren)}(\Sigma;\mu_{\rm IR}),
\label{eq:W-phase-odd-app-rew}
\end{equation}
A small-$\Sigma$ expansion of \eqref{eq:Isin-ren-def-rew}, whose full derivation is given in App.~\ref{app:small-expansion}, gives
\begin{equation}
I_{\sin}^{\rm (ren)}(\Sigma;\mu_{\rm IR})
=\frac{\kappa\,\Sigma}{4\pi^2}\ln\!\frac{\kappa}{2\pi\,\mu_{\rm IR}}
-\frac{7\zeta(3)}{48\pi^4}\,\kappa^3\Sigma^3+\mathcal{O}(\Sigma^5),
\end{equation}
which shows explicitly that, once the universal IR logarithm is subtracted, the odd channel is finite and analytic at short distances.

\subsection{\texorpdfstring{Regularity and singularity structure of the $\kappa\gamma$ vacuum}{}}
\label{subsec:singularity_kappagamma}

We now use the explicit expressions obtained above to clarify both the short--distance behavior and the global regularity of the $\kappa\gamma$ vacuum. The goal is to make precise two points: first, the state has the same ultraviolet singularity as the Minkowski vacuum (hence it is Hadamard); second, no additional divergences appear at finite $(u,v)$, in particular none are localized on the null lines that play the role of Rindler horizons.

The Wightman function splits into right-- and left--moving sectors in the usual way,
\begin{equation}
W_{\kappa\gamma}(x,x')
= W_{\kappa\gamma}^{\rm RTW}(u,u') + W_{\kappa\gamma}^{\rm LTW}(v,v'),
\end{equation}
with the right--moving contribution admitting the further decomposition
\begin{equation}
W_{\kappa\gamma}^{\rm RTW}(u,u')
= W^{\rm RTW}_{\rm th}(u-u') + W^{\rm RTW}_{\rm ph}(u+u';\gamma),
\label{eq:Wkappagamma_chiral-rew}
\end{equation}
and an analogous left--moving term obtained by $u\to v$, $u'\to v'$. The thermal piece is given by Eq.~\eqref{eq:Wth-RTW-app},
\begin{equation}
W^{\rm RTW}_{\rm th}(\Delta u)
= -\frac{1}{4\pi}\ln\!\Big[\mu^2\,\frac{2}{\kappa}\,
\sinh\!\Big(\frac{\kappa}{2}\,(\Delta u-i\epsilon)\Big)\Big],
\qquad \Delta u = u-u',
\label{eq:W_th_def-rew}
\end{equation}
and is precisely the standard chiral KMS correlator at temperature $T=\kappa/(2\pi)$. The phase--sensitive contribution depends instead on the sum coordinate $\Sigma=u+u'$:
\begin{equation}
W^{\rm RTW}_{\rm ph}(\Sigma;\gamma)
= W^{\rm RTW}_{\rm ph,even}(\Sigma;\gamma)
 +W^{\rm RTW}_{\rm ph,odd}(\Sigma;\gamma),
\qquad
\Sigma=u+u',
\end{equation}
where the explicit even and odd pieces are
\begin{align}
W^{\rm RTW}_{\rm ph,even}(\Sigma;\gamma)
&=-\frac{\cos(2\gamma)}{4\pi}\,
\ln\!\cosh\!\Big(\frac{\kappa\Sigma}{2}\Big),
\\[0.5ex]
W^{\rm RTW}_{\rm ph,odd}(\Sigma;\gamma)
&=\sin(2\gamma)\,I_{\sin}^{\rm (ren)}(\Sigma;\mu_{\rm IR}),
\end{align}
with $I_{\sin}^{\rm (ren)}$ defined in Eq.~\eqref{eq:Isin-ren-def-rew} and $\mu_{\rm IR}$ encoding the choice of infrared prescription.

\subsubsection*{Hadamard property and coincidence limit}

The Hadamard property is determined entirely by the behavior of $W_{\kappa\gamma}$ as the two points approach one another. For comparison, we recall that for a massless scalar field in $(1{+}1)$ dimensions the Minkowski vacuum Wightman function can be written as
\begin{equation}
W_{\rm M}(x,x')
= -\frac{1}{4\pi}\ln\!\big[\mu^2 (u-u'-i\epsilon)\big]
  -\frac{1}{4\pi}\ln\!\big[\mu^2 (v-v'-i\epsilon)\big],
\label{eq:W_Mink-rew}
\end{equation}
up to an additive state-independent constant.

The short--distance structure of the thermal term follows directly from expanding the hyperbolic sine in \eqref{eq:W_th_def-rew}. In the limit $\Delta u\to0$ one finds
\begin{equation}
W^{\rm RTW}_{\rm th}(\Delta u)
= -\frac{1}{4\pi}\left\{
    \ln\!\big[\mu^2(\Delta u-i\epsilon)\big]
   +\frac{\kappa^2}{24}(\Delta u)^2
   +O\big((\Delta u)^4\big)
  \right\},
\qquad \Delta u\to0.
\label{eq:W_th_shortdistance-rew}
\end{equation}
The leading logarithmic divergence is therefore identical to the right--moving Minkowski singularity in \eqref{eq:W_Mink-rew}; the parameter $\kappa$ enters only through smooth corrections that begin at order $(\Delta u)^2$.

Next, we examine the phase-sensitive terms. These depend on $\Sigma=u+u'$ rather than the separation $\Delta u$, and hence they remain finite in the coincidence limit at fixed midpoint. Using the explicit expressions derived above, we can verify smoothness directly. For the even part,
\begin{equation}
-\frac{\cos(2\gamma)}{4\pi}
\ln\!\cosh\!\Big(\frac{\kappa\Sigma}{2}\Big)
= -\frac{\cos(2\gamma)}{4\pi}
  \left[
    \frac{\kappa^2}{8}\Sigma^2
    +O\big(\Sigma^4\big)
  \right],
\label{eq:W_ph_even_shortdistance-rew}
\end{equation}
which is analytic at $\Sigma=0$. For the odd part, the renormalized definition \eqref{eq:Isin-ren-def-rew} implies the small-$\Sigma$ expansion
\begin{equation}
I_{\sin}^{\rm (ren)}(\Sigma;\mu_{\rm IR})
= \frac{\kappa}{4\pi^2}\,\Sigma\,
    \ln\!\Big(\frac{\kappa}{2\pi\mu_{\rm IR}}\Big)
  -\frac{7\zeta(3)}{48\pi^4}\,\kappa^3\Sigma^3
  +O(\kappa^5\Sigma^5),
\label{eq:I_sin_shortdistance-rew}
\end{equation}
which is likewise analytic for real $\Sigma$. Thus both $W^{\rm RTW}_{\rm ph,even}$ and $W^{\rm RTW}_{\rm ph,odd}$ are smooth functions of $(u,u')$, and the same conclusions hold for the left--moving sector as functions of $(v,v')$.

Putting these pieces together, we may express the full two-point function as
\begin{equation}
W_{\kappa\gamma}(x,x')
= W_{\rm M}(x,x') + F_{\kappa\gamma}(x,x'),
\label{eq:W_diff_smooth-rew}
\end{equation}
where $F_{\kappa\gamma}(x,x')$ is smooth in its arguments (in particular, in a neighborhood of the coincidence limit $x=x'$). All dependence on $\kappa$ and $\gamma$ beyond the universal Minkowski short-distance singularity is therefore carried by this smooth remainder. It follows that the $\kappa\gamma$ vacuum is a Hadamard state: its ultraviolet singularity structure coincides with that of the Minkowski vacuum, so local observables defined by point splitting are rendered finite once the Minkowski subtraction is performed.

Finally, the explicit expressions for the phase-sensitive channel involve only $\ln\cosh(\kappa\Sigma/2)$ and the renormalized odd function $I_{\sin}^{\rm (ren)}(\Sigma;\mu_{\rm IR})$, both of which are finite for all finite $\Sigma$. The thermal terms depend only on $\Delta u$ and $\Delta v$ and are likewise finite away from coincidence. Consequently the smooth remainder $F_{\kappa\gamma}(x,x')$ extends across the null lines $u=0$ and $v=0$ without developing any new singularities. In this sense, the $\kappa\gamma$ vacuum is not only locally Hadamard but also globally regular on Minkowski spacetime, with no additional divergences supported on the Rindler horizons.

\section{Discussion: The Interpretation of  \texorpdfstring{$\kappa\gamma$}{kappa-gamma} Vacuum}
\label{sec:discussion}

\subsection{The \texorpdfstring{$\kappa\gamma$}{kappa-gamma} Vacuum as a Continuous-Mode Phase-Squeezed State}

The phase parameter $\gamma$ extends the real squeezing of the $\kappa$-plane-wave vacuum into a genuinely phase-sensitive, continuous-mode squeezed state. The key point is that $\kappa$ fixes the frequency-dependent squeezing magnitude (and thus the thermal scale $T=\kappa/(2\pi)$), whereas $\gamma$ fixes a global squeeze angle, i.e., the orientation in phase space. In this subsection we make this structure explicit by relating $\ket{0_{\kappa,\gamma}}$ to the Minkowski vacuum $\ket{0_M}$, and we then use the same viewpoint to interpret the Bogoliubov maps between different $\kappa\gamma$ vacua.

The clearest physical reading of $\ket{0_{\kappa,\gamma}}$ follows from its relation to $\ket{0_M}$. The Bogoliubov transformation in Eq.~\eqref{eq:kg_mink_transf} connects the corresponding operator bases. Imposing the defining condition
\begin{equation}
{\cal A}_{\Lambda, \kappa, \gamma} \ket{0_{\kappa,\gamma}} = 0\,,
\end{equation}
and expressing ${\cal A}_{\Lambda,\kappa,\gamma}$ in terms of Minkowski operators gives
\begin{equation}
\frac{1}{\sqrt{2\sinh\!\big(\tfrac{\pi\Lambda}{\kappa}\big)}}
\left( e^{\frac{\pi\Lambda}{2\kappa}}e^{-i\gamma} \, a_\Lambda
      - e^{-\frac{\pi\Lambda}{2\kappa}}e^{i\gamma} \, a^\dagger_\Lambda \right)
\ket{0_{\kappa,\gamma}} = 0\,.
\end{equation}
Rearranging yields the mode-by-mode constraint
\begin{equation}
\left(a_{\Lambda} - e^{-\frac{\pi \Lambda}{\kappa}} e^{2i\gamma} a^{\dagger}_{\Lambda} \right)
\ket{0_{\kappa,\gamma}} = 0\,.
\label{eq:squeezing_condition_revised}
\end{equation}
This is precisely the defining relation of a squeezed vacuum with a \emph{complex} squeezing parameter~\cite{Scully_Zubairy1997, Agarwal2012}. In particular, the complex quantity
\begin{equation}
\zeta(\Lambda, \gamma) = e^{-\frac{\pi \Lambda}{\kappa}} e^{2i\gamma}
\end{equation}
encodes both the squeezing magnitude and the squeezing phase at frequency $\Lambda$.

Moreover, it is convenient to parameterize this relation in the standard single-mode language. A unitary squeezing operator acting on $\ket{0_M}$ may be written as
\begin{equation}
S(\xi)=\exp\!\Big\{\frac{1}{2}\big(\xi^* a^2 - \xi (a^\dagger)^2\big)\Big\},
\end{equation}
and a general single-mode squeezed state is described by $\xi = r e^{i\phi}$ and satisfies (see, for instance, Ref.~\cite{Azizi2025Unitary_TFD})
\begin{equation}
a \ket{\psi} = -\,e^{i\phi}\tanh r \; a^\dagger \ket{\psi}\,.
\end{equation}
Comparing this with Eq.~\eqref{eq:squeezing_condition_revised} gives the identifications
\begin{equation}
\tanh r(\Lambda) = e^{-\frac{\pi\Lambda}{\kappa}},\qquad
e^{i\phi(\Lambda)}=-\,e^{2i\gamma}
\quad\Longleftrightarrow\quad
\phi(\Lambda) \equiv 2\gamma+\pi \pmod{2\pi}\,.
\end{equation}
Equivalently, the squeezing magnitude and squeezing parameter can be written as
\begin{equation}
r(\Lambda)=\tanh^{-1}\!\Big(e^{-\frac{\pi\Lambda}{\kappa}}\Big)
=\frac{1}{2}\ln\!\coth\!\Big(\frac{\pi\Lambda}{2\kappa}\Big),
\end{equation}
and
\begin{equation}
\xi(\Lambda,\gamma)
= r(\Lambda)e^{i\phi(\Lambda)}
= -\,r(\Lambda)\,e^{2i\gamma}
= -\,\frac{1}{2}e^{2i\gamma}\ln\!\coth\!\Big(\frac{\pi\Lambda}{2\kappa}\Big).
\end{equation}
Accordingly, the $\kappa\gamma$ vacuum can be viewed as a continuous tensor product of single-mode squeezed vacua, with a frequency-dependent squeezing magnitude set by $\kappa$ and a uniform squeezing phase controlled by $\gamma$.

Furthermore, the same identification yields a compact formal expression for the state in terms of a continuous-mode squeezing operator acting on $\ket{0_M}$:
\begin{align}
\ket{0_{\kappa,\gamma}}
&= \exp\Bigg\{ \frac{1}{2} \int_0^\infty d\Lambda\, \Big( \xi^*(\Lambda,\gamma)\, a_{\Lambda}^2 \;-\; \xi(\Lambda,\gamma)\, (a^{\dagger}_{\Lambda})^2 \Big) \Bigg\} \ket{0_M},
\label{eq:squeezed_state_unitary_revised} \\
\text{where}\qquad
\xi(\Lambda,\gamma) &= -\frac{1}{2} e^{2i\gamma} \ln\!\left[ \coth\!\left( \frac{\pi \Lambda}{2\kappa} \right) \right]. \nn
\end{align}
Because $r(\Lambda)\to\infty$ as $\Lambda\to0^+$, this continuous-mode representation should be interpreted with the same infrared prescription used elsewhere in the paper (e.g., an IR scale $\mu_{\rm IR}$ or a finite-volume cutoff), so that the squeezing map is defined at the regulated level. With this understood, $\ket{0_{\kappa,\gamma}}$ is a continuous-mode phase-squeezed Gaussian state: $\kappa$ fixes the \emph{amount} of squeezing mode by mode, while $\gamma$ fixes the \emph{axis} in phase space.

Finally, the same squeezed-state viewpoint organizes the transformation between two arbitrary vacua $\ket{0_{\kappa,\gamma}}$ and $\ket{0_{\kappa',\gamma'}}$. Using Eq.~\eqref{eq:bogoliubov_kg_kg}, the defining condition ${\cal A}_{\Lambda,\kappa',\gamma'} \ket{0_{\kappa',\gamma'}} = 0$ can be expressed in the $(\kappa,\gamma)$ basis as
\begin{equation}
\left({\cal A}_{\Lambda,\kappa,\gamma} - \eta_{\Lambda} \, {\cal A}^{\dagger}_{\Lambda,\kappa,\gamma} \right)
\ket{0_{\kappa',\gamma'}} = 0\,.
\label{Kappa.g.PW.vs.Kappa.g.PW.vac1}
\end{equation}
The complex squeezing parameter is the ratio of Bogoliubov coefficients,
\begin{equation}
\eta_{\Lambda} =
\frac{\sinh \!\left( \alpha + i\Delta \right)}{\sinh \!\left( \beta - i\Delta \right)}\,, \qquad
\alpha = \frac{\pi \Lambda}{2} \!\left( \frac{1}{\kappa} - \frac{1}{\kappa'} \right),\quad
\beta = \frac{\pi \Lambda}{2} \!\left( \frac{1}{\kappa} + \frac{1}{\kappa'} \right),\quad
\Delta = \gamma' - \gamma.
\label{eta_gamma}
\end{equation}
Using $|\sinh(x+i\Delta)|^2 = \sinh^2 x + \sin^2\Delta$, one finds
\[
|\eta_\Lambda|^2=\frac{\sinh^2\alpha+\sin^2\Delta}{\sinh^2\beta+\sin^2\Delta}<1 \quad\text{for}\quad \kappa,\kappa'>0,
\]
since $\beta>\lvert\alpha\rvert$ whenever $\kappa,\kappa'>0$. Thus the map defines a bona fide squeeze. Writing
\[
\eta_\Lambda=\tanh r_\Lambda\,e^{i(\phi_\Lambda+\pi)},\qquad \xi_\Lambda=r_\Lambda e^{i \phi_\Lambda},
\]
it follows that
\[
r_\Lambda=\tanh^{-1}\!\big(|\eta_\Lambda|\big),\qquad \phi_\Lambda=\arg(\eta_\Lambda)-\pi.
\]
With these definitions, the state $\ket{0_{\kappa',\gamma'}}$ is obtained from $\ket{0_{\kappa,\gamma}}$ by a continuous-mode squeeze in either of the two equivalent forms
\begin{align}
\ket{0_{\kappa',\gamma'}}
&= \frac{1}{\sqrt{Z}}
\exp\!\left\{ \frac{1}{2} \int_0^\infty d\Lambda \, \eta_\Lambda \, \big({\cal A}^{\dagger}_{\Lambda, \kappa, \gamma}\big)^2 \right\}
\ket{0_{\kappa,\gamma}} \nn\\
&= \exp \!\left\{ \frac{1}{2} \int_0^\infty d\Lambda \, \tanh^{-1}\!\big(|\eta_\Lambda|\big)\,
\left( e^{-i \phi_{\Lambda}}
\big({\cal A}_{\Lambda, \kappa, \gamma}\big)^2
-  e^{i \phi_{\Lambda}}\,
\big({\cal A}^{\dagger}_{\Lambda, \kappa, \gamma}\big)^2   \right) \right\}
\ket{0_{\kappa,\gamma}}.
\end{align}
In this sense, the family of $\kappa\gamma$ vacua is connected by continuous-mode squeezing transformations generated by quadratic operators in the corresponding annihilation and creation operators.

Moreover, the Minkowski relations are recovered as a special case of the same general map. Taking the inverse transformation so that $(\kappa',\gamma')$ corresponds to the Minkowski frame, i.e.\ $\kappa'\to 0$, $\gamma'=0$, one obtains
\begin{equation}
 \left(a_{\Lambda} - \eta'_{\Lambda} \, a^{\dagger}_{\Lambda} \right) \ket{0_{\kappa,\gamma}} = 0,\qquad
 \eta'_{\Lambda}=\lim_{\substack{\kappa' \to 0 \\ \gamma' = 0}}
 \frac{ \sinh \!\left[ \frac{\pi \Lambda}{2} \!\left( \frac{1}{\kappa'} - \frac{1}{\kappa} \right) + i(\gamma - \gamma') \right] }{ \sinh \!\left[ \frac{\pi \Lambda}{2} \!\left( \frac{1}{\kappa'} + \frac{1}{\kappa} \right) - i(\gamma - \gamma') \right] }
 = e^{-\frac{\pi\Lambda}{\kappa}} e^{2i\gamma},
\end{equation}
which reproduces Eq.~\eqref{eq:squeezing_condition_revised}. Conversely, setting the state in Eq.~\eqref{Kappa.g.PW.vs.Kappa.g.PW.vac1} to $\ket{0_M}$ via the same limit gives
\begin{equation}
    \left({\cal A}_{\Lambda,\kappa,\gamma} - \eta_{\Lambda}^{\text{(Mink)}} \, {\cal A}^{\dagger}_{\Lambda,\kappa,\gamma} \right) \ket{0_{M}} = 0,
\end{equation}
where
\begin{equation}
    \eta_{\Lambda}^{\text{(Mink)}}=\lim_{\substack{\kappa' \to 0 \\ \gamma' = 0}}
    \frac{\sinh \!\left[ \frac{\pi \Lambda}{2} \!\left( \frac{1}{\kappa} - \frac{1}{\kappa'} \right) + i(\gamma' - \gamma) \right]}{\sinh \!\left[ \frac{\pi \Lambda}{2} \!\left( \frac{1}{\kappa} + \frac{1}{\kappa'} \right) - i(\gamma' - \gamma) \right]}
    = -\,e^{-\frac{\pi\Lambda}{\kappa}}\,e^{-2i\gamma}.
\end{equation}
Hence
\begin{equation}
    \left({\cal A}_{\Lambda,\kappa,\gamma} + e^{-\frac{\pi\Lambda}{\kappa}}\,e^{-2i\gamma} \, {\cal A}^{\dagger}_{\Lambda,\kappa,\gamma} \right) \ket{0_{M}} = 0,
\end{equation}
and the two limits are related by
\[
\eta_{\Lambda}^{\text{(Mink)}} = -\,e^{-2i\gamma}\,\eta'_\Lambda,
\]
as expected from inverse squeezing in different operator bases.

\subsection{\texorpdfstring{Physical meaning of $\kappa$ and $\gamma$: a modulated accelerating-mirror origin}{}}

The parameters $\kappa$ and $\gamma$ admit a direct dynamical interpretation: they can be realized as independently tunable controls in a modulated accelerating-mirror setup~\cite{Azizi2025ModMirror}. In that realization, the accelerating trajectory fixes the \emph{Planckian weights} of the Bogoliubov mixing, thereby setting the effective temperature scale $T=\kappa/(2\pi)$, while a weak, chiral, frequency-diagonal boundary modulation imprints an overall \emph{phase} on the mixing coefficients~\cite{Azizi2025ModMirror}. Operationally, this boundary modulation can be implemented as a time-dependent Robin (impedance) law on the mirror worldline; its effect is to rotate the squeeze angle (controlled by $\gamma$) without changing the Planckian modulus determined by $\kappa$, to leading order~\cite{Azizi2025ModMirror}.

This separation shows up transparently at future null infinity: the resulting two-point function decomposes into a stationary thermal contribution and a phase-sensitive, non-stationary contribution~\cite{Azizi2025ModMirror}. Inertial Unruh--DeWitt detectors therefore register an exact Planck law governed by $\kappa$, whereas uniformly accelerated detectors can access $\gamma$ through interference effects and may exhibit mode-selective suppression when the phase-sensitive channel contributes coherently~\cite{Azizi2025ModMirror}. In this sense, $\kappa$ fixes the strength of the thermal/squeezing weights, and $\gamma$ fixes the orientation of the squeezing in phase space.

Two technical remarks align this interpretation with the calculations in the present manuscript. First, the mapping to an ideal $\kappa\gamma$ vacuum is formulated asymptotically (on $\mathscr{I}^\pm$), where the chiral sector is the natural language for the outgoing radiation~\cite{Azizi2025ModMirror}. Second, if the boundary modulation produces a weak spectral dependence $\gamma_{\rm eff}(\Lambda)$ across the thermally relevant band, then the phase-sensitive signatures become spectrally structured; when $\gamma_{\rm eff}(\Lambda)\approx \gamma$ is effectively flat, the constant-$\gamma$ description used here is recovered~\cite{Azizi2025ModMirror}.

Finally, the infrared bookkeeping introduced in Sec.~\ref{subsec:phase-Sigma-IR} has a concrete meaning in this dynamical picture: the odd, phase-sensitive channel is defined only up to an IR subtraction (encoded by $\mu_{\rm IR}$), reflecting that an IR-sensitive linear-in-$\Sigma$ contribution corresponds to a phase-convention choice, while physically diagnostic quantities are contained in IR-stable, phase-sensitive differences and derivatives.

\subsection{Visualizing the Mode Structure and Its Consequences}
\begin{figure}[t]
    \centering
    \includegraphics[width=0.85\textwidth]{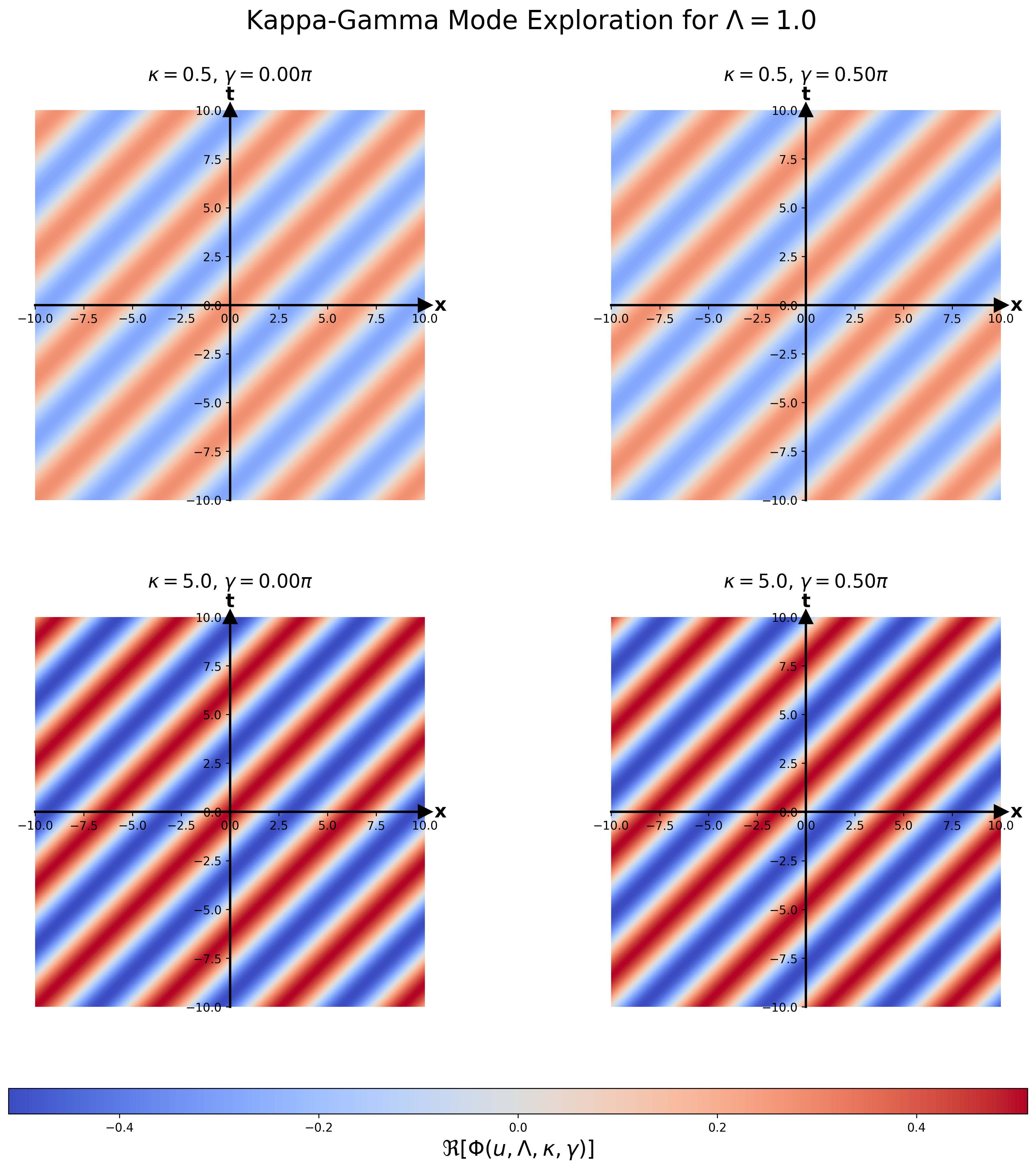}
    \caption{Structure of the $\kappa\gamma$-mode for a representative frequency of $\Lambda=1.0$. The panels show the real part of the wave function for different combinations of the deformation parameter $\kappa$ and the phase factor $\gamma$. The top row corresponds to a smaller deformation ($\kappa=0.5$), while the bottom row shows a larger deformation ($\kappa=5.0$). The columns represent phase shifts of $\gamma=0$ (left) and $\gamma=\pi/2$ (right).}
    \label{fig:mode_structure}
\end{figure}

\begin{figure}[t]
    \centering
    \includegraphics[width=\textwidth]{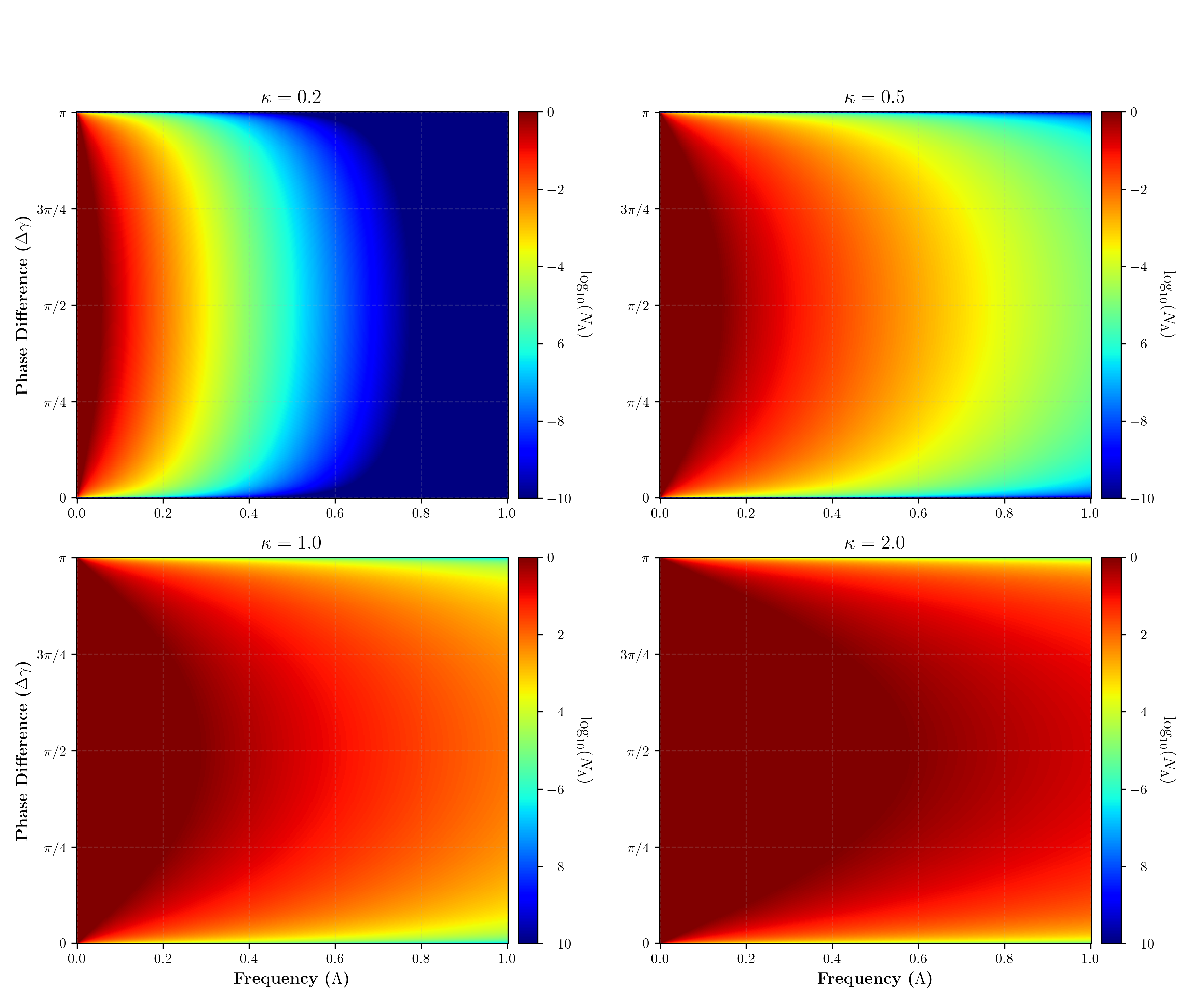}
    \caption{Phase-induced particle creation for different vacua. Each panel shows the particle number spectrum $N_\Lambda$ on a logarithmic scale as a function of frequency $\Lambda$ and phase difference $\Delta\gamma$ for a fixed value of $\kappa$. For small $\kappa$, particle creation is suppressed at all but the lowest frequencies; as $\kappa$ increases, the effect strengthens and extends to higher $\Lambda$, with the maximum always at $\Delta\gamma=\pi/2$.}
    \label{fig:kappa_grid_particle_creation}
\end{figure}

To build intuition for the algebraic results, we can visualize both the field modes themselves and the particle creation that results from their structure. The parameters $\Lambda$, $\kappa$, and $\gamma$ each play a distinct physical role, which is illustrated in Figs.~\ref{fig:mode_structure} and \ref{fig:kappa_grid_particle_creation}.

First, we examine the structure of the $\kappa\gamma$-mode function, $\Phi(u,\Lambda,\kappa,\gamma)$, shown in Fig.~\ref{fig:mode_structure} for a representative frequency of $\Lambda=1.0$. The plot reveals the physical meaning of the parameters $\kappa$ and $\gamma$:
\begin{enumerate}
    \item The deformation parameter $\kappa$ governs the magnitude of the deviation from a pure plane wave. For small $\kappa$ (top row), the mode is nearly a simple, right-moving Minkowski plane wave. For larger $\kappa$ (bottom row), the mode deviates significantly due to a more balanced mixing of positive- and negative-frequency components, leading to a pronounced interference pattern.
    \item The phase parameter $\gamma$ introduces a spatial shift in this interference pattern. This is clearly seen by comparing the left ($\gamma=0$) and right ($\gamma=\pi/2$) columns, where the crests and troughs of the interference fringes are translated relative to each other.
\end{enumerate}

The direct physical consequence of this phase-controllable mode structure is the phenomenon of phase-induced particle creation, visualized in Fig.~\ref{fig:kappa_grid_particle_creation}. This figure shows the particle number spectrum $N_\Lambda$ as a function of the phase difference $\Delta\gamma$ and the absolute frequency $\Lambda$ for several different values of the underlying vacuum parameter $\kappa$. Across all panels, the spectrum is maximized for a phase difference of $\Delta\gamma = \pi/2$ and vanishes when the frames are in phase ($\Delta\gamma=0$), in agreement with Eq.~\eqref{eq:particle_number}. The crucial role of $\kappa$ as a control parameter is also evident: for small $\kappa$ (e.g., top-left, $\kappa=0.2$), particle creation is strongly suppressed except in the deep infrared, where the $\Lambda\to0$ enhancement is visible. As $\kappa$ increases, the overall magnitude of $N_\Lambda$ grows and the enhancement extends to higher frequencies, so that phase-induced particle creation becomes both stronger and more broadband. Thus, while a larger $\kappa$ corresponds to stronger interference in the mode structure, it also leads to a more pronounced particle creation effect for any fixed phase mismatch.

\section{Conclusion} \label{sec:conclusion}

In this work we extended the $\kappa$ plane-wave construction of Ref.~\cite{Azizi2025KappaPW} by introducing the $\kappa\gamma$ plane-wave modes, characterized by a deformation scale $\kappa$ together with an independent phase parameter $\gamma$. This enlargement produces a broader family of vacua and admits a boundary-driven realization in Carlitz--Willey type moving-mirror settings: the accelerating trajectory fixes the Planckian modulus kinematically, while a weak, chiral, approximately frequency-diagonal Robin (impedance) modulation rotates the squeezing axis without modifying those weights at leading order~\cite{Azizi2025ModMirror}.

A central outcome is that a \emph{relative} phase mismatch between two $\kappa\gamma$ bases produces genuine Bogoliubov mixing. As a result, the vacuum associated with one phase does not, in general, appear empty when expanded in the other basis. This effect is quantified by the particle number density
\begin{equation}
N_\Lambda(\gamma',\gamma)
=\frac{\sin^2(\gamma'-\gamma)}{\sinh^2(\pi\Lambda/\kappa)}\,,
\end{equation}
which isolates the physical role of the phase difference $\Delta\gamma=\gamma'-\gamma$. In the mirror picture, $\kappa$ fixes the Planckian weights determined by the background trajectory, whereas $\gamma$ is imprinted as a controlled rotation of the squeeze angle generated by the boundary drive~\cite{Azizi2025ModMirror}.

We also established that the $\kappa\gamma$ vacuum $\ket{0_{\kappa,\gamma}}$ is a continuous-mode \emph{phase-squeezed} Gaussian state relative to the Minkowski vacuum. The continuous squeezing structure already present in the $\kappa$ plane-wave vacuum~\cite{Azizi2025KappaPW} persists, while the new parameter $\gamma$ selects a global squeezing axis in phase space. In this way, the real squeezing of the $\kappa$ plane-wave family is promoted to a genuinely phase-sensitive family with identical Planck weights but phase-dependent correlations~\cite{Azizi2025ModMirror}.

At the level of two-point functions, the Wightman function makes the regularity of the state explicit. Its thermal part reproduces the standard Minkowski short-distance singularity with only smooth $\kappa$-dependent corrections, while the phase-sensitive channel depends on the sum coordinate and is analytic at coincidence. After the standard infrared renormalization of the odd channel, the full $\kappa\gamma$ correlator differs from the Minkowski correlator by a smooth function. Consequently the $\kappa\gamma$ vacuum is Hadamard and exhibits no ultraviolet singularities beyond those of the Minkowski vacuum, with no additional divergences localized on the Rindler horizons.

It is also useful to emphasize how this plane-wave sector complements the earlier $\kappa$-Rindler sector. The $\kappa$-Rindler vacua are boost-stationary and use $\kappa$ to deform the effective thermality perceived by uniformly accelerated detectors~\cite{Azizi2025Tunable}, whereas the $\kappa$ plane-wave (and hence $\kappa\gamma$) quantization is translation-invariant and is not continuously connected to the Rindler vacuum in any limit~\cite{Azizi2025KappaPW}.

Finally, we derived the generalized \textit{Mother of All Bogoliubov Transformations}, including the extension by the phase parameter $\gamma$. This master map provides a unified framework that contains the mode decompositions used throughout the paper and recovers earlier constructions in the appropriate limits, while keeping the role of $\gamma$ explicit at every step. As in any continuum squeezing construction, the associated unitaries are understood with an infrared regulator (finite volume, wave packets, or smooth switching); once this is implemented, physical observables remain well behaved under regulator removal.

\section*{Acknowledgments}

I am grateful to Marlan Scully, Anatoly Svidzinsky, and Bill Unruh for illuminating discussions. I would also like to extend special thanks to an anonymous referee for an insightful question on a related work \cite{Azizi2025KappaPW} concerning the uniqueness of the mode coefficients; that question was a primary motivation for exploring the physical role of the relative phase parameter $\gamma$ in this paper. This work was supported by the Robert A. Welch Foundation (Grant No. A-1261).

\appendix

\section{Notations} \label{app:notation}

Throughout this paper, we use the notation $\Lambda \in \mathbb{R}^+$ to denote the frequency label for the $\kappa\gamma$ plane-wave modes. The operator ${\cal A}_{\Lambda, \kappa, \gamma}$ represents the annihilation operator associated with such a mode. By contrast, $\Omega \in \mathbb{R}$ denotes the frequency parameter used for $\kappa$-Rindler modes, following the convention introduced in Ref.~\cite{Azizi2023JHEP}. The corresponding annihilation operators for the $\kappa$-Rindler modes are denoted by ${\cal B}_{\Omega, \kappa}$.

\section{Derivation: \texorpdfstring{$\kappa\gamma$}{kappa-gamma}-plane wave in terms of \texorpdfstring{$\kappa'$}{kappa'}-Rindler operators}
\label{app:derivation_kg_in_kr}

This appendix provides a detailed derivation of the Bogoliubov transformation that expresses the $\kappa\gamma$ plane-wave operators ${\cal A}_{\Lambda, \kappa, \gamma}$ in terms of the $\kappa'$-Rindler operators ${\cal B}_{\Omega, \kappa'}$.

Our strategy is as follows:
\begin{enumerate}
\item We equate the scalar field operator $\Phi(u)$ expanded in the $\kappa\gamma$ plane-wave basis with its expansion in the $\kappa'$-Rindler basis.
\item We perform Fourier projections on this identity by integrating against the kernels $e^{\pm i\Lambda_0 u}$, thereby generating a coupled system of linear algebraic equations.
\item We resolve this system to isolate the annihilation operator ${\cal A}_{\Lambda_0, \kappa, \gamma}$.
\end{enumerate}

\subsection{Fourier projection of the field expansions}
\label{app:fourier-projection}

First, we project the $\kappa\gamma$ plane-wave expansion. Applying the Fourier transform to the field $\Phi(u)$ for a fixed frequency $\Lambda_0>0$ yields
\begin{align}
\int_{-\infty}^{+\infty} du \, e^{i\Lambda_0 u} \Phi(u)
&= \frac{2\pi}{\sqrt{{\cal N}_{\Lambda_0,\kappa}}}
\Bigg\{ e^{\frac{\pi\Lambda_0}{2\kappa} + i\gamma} {\cal A}_{\Lambda_0,\kappa,\gamma}
      + e^{-\frac{\pi\Lambda_0}{2\kappa} + i\gamma} {\cal A}^\dagger_{\Lambda_0,\kappa,\gamma} \Bigg\},
\label{eq:fourier_kg_1}\\
\int_{-\infty}^{+\infty} du \, e^{-i\Lambda_0 u} \Phi(u)
&= \frac{2\pi}{\sqrt{{\cal N}_{\Lambda_0,\kappa}}}
\Bigg\{ e^{-\frac{\pi\Lambda_0}{2\kappa} - i\gamma} {\cal A}_{\Lambda_0,\kappa,\gamma}
      + e^{\frac{\pi\Lambda_0}{2\kappa} - i\gamma} {\cal A}^\dagger_{\Lambda_0,\kappa,\gamma} \Bigg\}.
\label{eq:fourier_kg_2}
\end{align}

Next, we apply the same Fourier projections to the $\kappa'$-Rindler expansion of $\Phi(u)$. Writing
\begin{align}
{\cal M}_{\Lambda_0}
\equiv \int_{-\infty}^{+\infty} du\, e^{i\Lambda_0 u}\, \Phi(u)\bigg|_{\kappa'\text{-Rindler}},
\label{eq:M-def-app}
\end{align}
the evaluation reduces to the regulated integrals in Appendix~\ref{app:gamma-proj}. Concretely, the $\theta(u)$ and $\theta(-u)$ pieces produce the Fourier--Gamma factors \eqref{eq:int-pos-plus-app}--\eqref{eq:int-neg-minus-app}, which combine with the wedge weights $e^{\pm \frac{\pi\Omega}{2\kappa'}}$ from the $\kappa'$-Rindler mode functions. After assembling the two wedge contributions and absorbing convention-dependent overall phases, one finds
\begin{align}
{\cal M}_{\Lambda_0}
=&\int_{-\infty}^{+\infty} \frac{d\Omega \; i}{\sqrt{{\cal N}_{\Omega,\kappa'}}}
\Bigg\{
\left(
-\, e^{\frac{\pi \Omega}{2}} e^{\frac{\pi \Omega}{2\kappa'}}
+    e^{-\frac{\pi \Omega}{2}} e^{-\frac{\pi \Omega}{2\kappa'}}
\right)\Lambda_0^{-(1+i\Omega)} 
\Gamma(1+i\Omega)\, {\cal B}_{\Omega,\kappa'}\nn\\
&\phantom{\int_{-\infty}^{+\infty} \frac{d\Omega \; i}{\sqrt{{\cal N}_{\Omega,\kappa'}}}
\Bigg\{} 
+\left(
-\, e^{-\frac{\pi \Omega}{2}} e^{\frac{\pi \Omega}{2\kappa'}}
+    e^{\frac{\pi \Omega}{2}} e^{-\frac{\pi \Omega}{2\kappa'}}
\right)\Lambda_0^{-(1-i\Omega)}
\Gamma(1-i\Omega)\, {\cal B}^\dagger_{\Omega,\kappa'}
\Bigg\}.
\label{eq:M-Lambda0-final-app}
\end{align}
For later convenience, the two parenthetical factors can be regrouped as
\begin{align}
-\, e^{\frac{\pi \Omega}{2}} e^{\frac{\pi \Omega}{2\kappa'}}
+    e^{-\frac{\pi \Omega}{2}} e^{-\frac{\pi \Omega}{2\kappa'}}
&=-2\sinh\!\left(\frac{\pi\Omega}{2}\Big(\frac{1}{\kappa'}+1\Big)\right),\nn\\
-\, e^{-\frac{\pi \Omega}{2}} e^{\frac{\pi \Omega}{2\kappa'}}
+    e^{\frac{\pi \Omega}{2}} e^{-\frac{\pi \Omega}{2\kappa'}}
&=-2\sinh\!\left(\frac{\pi\Omega}{2}\Big(\frac{1}{\kappa'}-1\Big)\right).
\label{eq:M-parentheses-to-sinh-app}
\end{align}
Since $\Phi(u)$ is Hermitian, the second Fourier projection satisfies
\begin{equation}
{\cal M}^{\dagger}_{\Lambda_0}
=\left(\int_{-\infty}^{+\infty} du \, e^{i\Lambda_0 u} \Phi(u)\right)^\dagger
=\int_{-\infty}^{+\infty} du \, e^{-i\Lambda_0 u} \Phi(u),
\end{equation}
and we denote it by ${\cal M}^{\dagger}_{\Lambda_0}$.

\subsection{System of equations and solution}

Equating the Fourier projections in the two bases yields
\begin{align}
\frac{2\pi}{\sqrt{{\cal N}_{\Lambda_0,\kappa}}}
\Bigg\{ e^{\frac{\pi\Lambda_0}{2\kappa} + i\gamma} {\cal A}_{\Lambda_0,\kappa,\gamma}
      + e^{-\frac{\pi\Lambda_0}{2\kappa} + i\gamma} {\cal A}^\dagger_{\Lambda_0,\kappa,\gamma} \Bigg\}
&= {\cal M}_{\Lambda_0}, \\
\frac{2\pi}{\sqrt{{\cal N}_{\Lambda_0,\kappa}}}
\Bigg\{ e^{-\frac{\pi\Lambda_0}{2\kappa} - i\gamma} {\cal A}_{\Lambda_0,\kappa,\gamma}
      + e^{\frac{\pi\Lambda_0}{2\kappa} - i\gamma} {\cal A}^\dagger_{\Lambda_0,\kappa,\gamma} \Bigg\}
&= {\cal M}^{\dagger}_{\Lambda_0}.
\end{align}
In matrix form,
\begin{align}
\begin{pmatrix}
e^{\frac{\pi\Lambda_0}{2\kappa}}e^{i\gamma} & e^{-\frac{\pi\Lambda_0}{2\kappa}}e^{i\gamma} \\[0.2em]
e^{-\frac{\pi\Lambda_0}{2\kappa}}e^{-i\gamma} & e^{\frac{\pi\Lambda_0}{2\kappa}}e^{-i\gamma}
\end{pmatrix}
\begin{pmatrix}
{\cal A}_{\Lambda_0, \kappa, \gamma} \\
{\cal A}^\dagger_{\Lambda_0, \kappa, \gamma}
\end{pmatrix}
= \frac{\sqrt{{\cal N}_{\Lambda_0,\kappa}}}{2\pi}
\begin{pmatrix}
{\cal M}_{\Lambda_0} \\
{\cal M}^{\dagger}_{\Lambda_0}
\end{pmatrix}.
\end{align}
The determinant is
\begin{equation}
\det
= e^{\frac{\pi\Lambda_0}{\kappa}}-e^{-\frac{\pi\Lambda_0}{\kappa}}
=2\sinh\!\left(\frac{\pi\Lambda_0}{\kappa}\right),
\end{equation}
so the inverse system gives
\begin{align}
\begin{pmatrix}
{\cal A}_{\Lambda_0, \kappa, \gamma} \\
{\cal A}^\dagger_{\Lambda_0, \kappa, \gamma}
\end{pmatrix}
&=
\frac{1}{2\sinh\!\left(\frac{\pi\Lambda_0}{\kappa}\right)}
\begin{pmatrix}
e^{\frac{\pi\Lambda_0}{2\kappa}}e^{-i\gamma} 
& -e^{-\frac{\pi\Lambda_0}{2\kappa}}e^{i\gamma} \\
-e^{-\frac{\pi\Lambda_0}{2\kappa}}e^{-i\gamma}
& e^{\frac{\pi\Lambda_0}{2\kappa}}e^{i\gamma}
\end{pmatrix}
\frac{\sqrt{{\cal N}_{\Lambda_0,\kappa}}}{2\pi}
\begin{pmatrix}
{\cal M}_{\Lambda_0} \\
{\cal M}^{\dagger}_{\Lambda_0}
\end{pmatrix}.
\end{align}
Extracting the top component yields
\begin{align}
{\cal A}_{\Lambda_0,\kappa,\gamma}
= \frac{\sqrt{{\cal N}_{\Lambda_0,\kappa}}}{4\pi\sinh(\frac{\pi\Lambda_0}{\kappa})}
\left( e^{\frac{\pi\Lambda_0}{2\kappa}} e^{-i\gamma} {\cal M}_{\Lambda_0}
     - e^{-\frac{\pi\Lambda_0}{2\kappa}} e^{i\gamma} {\cal M}^{\dagger}_{\Lambda_0} \right).
\label{eq:A-in-terms-of-M-app}
\end{align}

\subsection{Final result}

Substituting the integral form of ${\cal M}_{\Lambda_0}$ and simplifying the hyperbolic combinations using \eqref{eq:M-parentheses-to-sinh-app}, we arrive at the final Bogoliubov transformation:
\begin{align}
&{\cal A}_{\Lambda,\kappa,\gamma}= \label{KGPW-in-KR_app}\\
& -\frac{i}{2 \pi} \int_{-\infty}^{\infty} d \Omega \,
\frac{1}{\sqrt{\Lambda \Omega \sinh(\frac{\pi \Lambda}{\kappa}) \sinh(\frac{\pi \Omega}{\kappa'})}}
 \nn\\
& \times \Bigg \{
\Lambda^{-i \Omega}\, \Gamma(1+i \Omega)
\Bigg[ e^{\frac{\pi \Lambda}{2 \kappa}-i\gamma} 
\sinh \left(\ft{\pi \Omega}{2}\Big(\frac{1}{\kappa'}+1\Big)\right) 
+ e^{-\frac{\pi \Lambda}{2 \kappa}+i\gamma}
\sinh \left(\ft{\pi \Omega}{2}\Big(\frac{1}{\kappa'}-1\Big)\right) \Bigg] {\cal B}_{\Omega,\kappa'} \nn\\
& \qquad +
\Lambda^{i \Omega}\, \Gamma(1-i \Omega)
\Bigg[ e^{\frac{\pi \Lambda}{2 \kappa}-i\gamma} 
\sinh \left(\ft{\pi \Omega}{2}\Big(\frac{1}{\kappa'}-1\Big)\right) 
+ e^{-\frac{\pi \Lambda}{2 \kappa}+i\gamma}
\sinh \left(\ft{\pi \Omega}{2}\Big(\frac{1}{\kappa'}+1\Big)\right) \Bigg] {\cal B}^{\dagger}_{\Omega,\kappa'}
\Bigg \}\,. \nn
\end{align}
Here the square-root factor is well-defined for $\Omega\neq 0$ since $\Omega\,\sinh(\pi\Omega/\kappa')>0$. This is the first of the two ``Mother of All Bogoliubov Transformations,'' generalized to include the phase parameter $\gamma$.

\section{Derivation: \texorpdfstring{$\kappa'$}{kappa'}-Rindler in terms of \texorpdfstring{$\kappa\gamma$}{kappa-gamma} plane-wave operators}
\label{app:derivation_kr_in_kg}

This appendix derives the Bogoliubov transformation expressing the $\kappa'$-Rindler
operators ${\cal B}_{\Omega,\kappa'}$ in terms of the $\kappa\gamma$ plane-wave
operators ${\cal A}_{\Lambda,\kappa,\gamma}$. The derivation is based on Mellin-type
projections of the field $\Phi(u)$ on the right and left wedges and on solving the
resulting $2\times2$ linear system.

We allow $\Omega\in\mathbb{R}$ throughout as a Mellin label. Since $\sinh(\pi\Omega/\kappa')$
is odd, the product $\Omega\,\sinh(\pi\Omega/\kappa')$ is strictly positive for $\Omega\neq0$,
so the combined square-root factors appearing below are real (equivalently, one may rewrite them
using $|\Omega|$). When expressing the inverse Bogoliubov map in a single formula valid for
$\Omega\in\mathbb{R}$, an overall $\operatorname{sgn}(\Omega)$ appears; this is equivalent to
working with a kernel written in terms of $|\Omega|$ and restricting to $\Omega>0$ in the usual
positive-frequency Rindler convention.

\subsection{Field expansions}

We start from the $\kappa\gamma$ plane-wave expansion
\begin{align}
\Phi(u) = \int_{0}^{\infty} \frac{d\Lambda}{\sqrt{8 \pi \Lambda \sinh(\frac{\pi \Lambda}{\kappa})}} \bigg\{ &
\Big(e^{\frac{\pi \Lambda}{2\kappa}} e^{i\gamma} e^{-i\Lambda u}
    + e^{-\frac{\pi \Lambda}{2\kappa}} e^{-i\gamma} e^{i\Lambda u}\Big) {\cal A}_{\Lambda,\kappa,\gamma} \nn \\
&+ \Big(e^{\frac{\pi \Lambda}{2\kappa}} e^{-i\gamma} e^{i\Lambda u}
    + e^{-\frac{\pi \Lambda}{2\kappa}} e^{i\gamma} e^{-i\Lambda u}\Big) {\cal A}^{\dagger}_{\Lambda,\kappa,\gamma} \bigg\}\,.
\label{eq:Phi-kg-exp-app}
\end{align}
The $\kappa'$-Rindler expansion is written piecewise across the right ($u<0$) and
left ($u>0$) wedges as
\begin{align}
\Phi(u) = &\;
\theta(-u) \int_{-\infty}^{\infty} \frac{d\Omega'}{\sqrt{8 \pi \Omega' \sinh(\frac{\pi \Omega'}{\kappa'})}}
\Big\{(-u)^{i\Omega'} e^{\frac{\pi \Omega'}{2\kappa'}} {\cal B}_{\Omega',\kappa'} + \text{h.c.}\Big\} \nn \\
&+ \theta(u) \int_{-\infty}^{\infty} \frac{d\Omega'}{\sqrt{8 \pi \Omega' \sinh(\frac{\pi \Omega'}{\kappa'})}}
\Big\{ u^{i\Omega'} e^{-\frac{\pi \Omega'}{2\kappa'}} {\cal B}_{\Omega',\kappa'} + \text{h.c.}\Big\}\,.
\label{eq:Phi-kr-exp-app}
\end{align}
Here ``h.c.'' denotes the Hermitian conjugate of the preceding term. This is a compact way of
writing the standard $\Omega'>0$ mode expansion; equivalently, one may restrict $\Omega'>0$ and
write the conjugate term explicitly, with no double counting implied.

\subsection{Projection onto Rindler modes}

To isolate ${\cal B}_{\Omega,\kappa'}$ we use weighted Mellin projections on each
wedge. The basic distributional identity is
\begin{equation}
\int_{0}^{\infty} du\,u^{\,i(\Omega+\Omega')-1}
=2\pi\,\delta(\Omega+\Omega'),
\label{eq:Mellin-delta-app}
\end{equation}
which follows by the change of variables $u=e^{x}$, so that the integral becomes
$\int_{-\infty}^{\infty}dx\,e^{i(\Omega+\Omega')x}$.

We obtain the following two equations for ${\cal B}_{-\Omega,\kappa'}$ and
${\cal B}^\dagger_{\Omega,\kappa'}$:
\begin{enumerate}
\item Projection onto the right wedge ($u<0$):
\begin{align}
\int_{-\infty}^0 du\, (-u)^{i\Omega-1} \Phi(u)
= \frac{2\pi}{\sqrt{8\pi\Omega\sinh(\frac{\pi\Omega}{\kappa'})}}
\left( e^{-\frac{\pi\Omega}{2\kappa'}}{\cal B}_{-\Omega,\kappa'} 
+ e^{\frac{\pi\Omega}{2\kappa'}} 
{\cal B}^\dagger_{\Omega,\kappa'} \right)\,.
\label{eq:proj-right-app}
\end{align}

\item Projection onto the left wedge ($u>0$):
\begin{align}
\int_0^\infty du\, u^{i\Omega-1} \Phi(u)
= \frac{2\pi}{\sqrt{8\pi\Omega\sinh(\frac{\pi\Omega}{\kappa'})}} 
\left( e^{\frac{\pi\Omega}{2\kappa'}}{\cal B}_{-\Omega,\kappa'} 
+ e^{-\frac{\pi\Omega}{2\kappa'}} {\cal B}^\dagger_{\Omega,\kappa'} \right)\,.
\label{eq:proj-left-app}
\end{align}
\end{enumerate}
Both relations use \eqref{eq:Mellin-delta-app}: the kernel $u^{i\Omega-1}$ paired
with $u^{i\Omega'}$ yields $\delta(\Omega+\Omega')$, while pairing with the h.c.\ term
$u^{-i\Omega'}$ yields $\delta(\Omega-\Omega')$.

\subsection{Solving the system of equations}

We now evaluate the same two wedge projections using the $\kappa\gamma$ expansion
\eqref{eq:Phi-kg-exp-app}. Upon inserting \eqref{eq:Phi-kg-exp-app} into the weighted
integrals \eqref{eq:proj-right-app}--\eqref{eq:proj-left-app}, the $u$-integrations reduce
to the regulated Mellin--Fourier integrals derived in Appendix~\ref{app:gamma}. This
produces $\Lambda$-integrals that we write schematically as
\begin{align}
\int_{-\infty}^0 du\, (-u)^{i\Omega-1} \Phi(u)
&= \int_0^\infty d\Lambda \Big(M_{\Omega\Lambda}\,{\cal A}_{\Lambda,\kappa,\gamma}
+ N_{\Omega\Lambda}\,{\cal A}^\dagger_{\Lambda,\kappa,\gamma}\Big),
\label{eq:proj-right-kernels-app}\\
\int_{0}^\infty du\, u^{i\Omega-1} \Phi(u)
&= \int_0^\infty d\Lambda \Big(K_{\Omega\Lambda}\,{\cal A}_{\Lambda,\kappa,\gamma}
+ P_{\Omega\Lambda}\,{\cal A}^\dagger_{\Lambda,\kappa,\gamma}\Big),
\label{eq:proj-left-kernels-app}
\end{align}
where the explicit kernel expressions follow directly from the Gamma integrals in
Appendix~\ref{app:gamma} (see also the collected forms in
Eqs.~\eqref{eq:Mkernel}--\eqref{eq:Pkernel}).

Equating \eqref{eq:proj-right-kernels-app}--\eqref{eq:proj-left-kernels-app} with the
corresponding Rindler-side projections \eqref{eq:proj-right-app}--\eqref{eq:proj-left-app}
yields a coupled $2\times2$ linear system for the pair
$\big({\cal B}_{-\Omega,\kappa'},{\cal B}^\dagger_{\Omega,\kappa'}\big)$. In matrix form,
\begin{align}
\begin{pmatrix}
e^{-\frac{\pi\Omega}{2\kappa'}} & e^{\frac{\pi\Omega}{2\kappa'}} \\
e^{\frac{\pi\Omega}{2\kappa'}} & e^{-\frac{\pi\Omega}{2\kappa'}}
\end{pmatrix}
\begin{pmatrix}
{\cal B}_{-\Omega,\kappa'} \\
{\cal B}^\dagger_{\Omega,\kappa'}
\end{pmatrix}
= \frac{\sqrt{8\pi\Omega\sinh(\frac{\pi\Omega}{\kappa'})}}{2\pi}
\int_0^{\infty} d\Lambda
\begin{pmatrix}
M_{\Omega\Lambda} & N_{\Omega\Lambda} \\
K_{\Omega\Lambda} & P_{\Omega\Lambda}
\end{pmatrix}
\begin{pmatrix}
{\cal A}_{\Lambda,\kappa,\gamma} \\
{\cal A}^\dagger_{\Lambda,\kappa,\gamma}
\end{pmatrix}.
\label{eq:matrix-system-app}
\end{align}
The determinant of the left matrix is
\begin{equation}
\det
=e^{-\frac{\pi\Omega}{\kappa'}}-e^{\frac{\pi\Omega}{\kappa'}}
=-2\sinh\!\Big(\frac{\pi\Omega}{\kappa'}\Big),
\end{equation}
which is nonzero for $\Omega\neq0$. The system is therefore invertible, and solving for
${\cal B}_{-\Omega,\kappa'}$ and then relabeling $\Omega\to-\Omega$ yields
${\cal B}_{\Omega,\kappa'}$ as a $\Lambda$-integral over
${\cal A}_{\Lambda,\kappa,\gamma}$ and ${\cal A}^\dagger_{\Lambda,\kappa,\gamma}$, with
all coefficients fixed by the kernel evaluation in Appendix~\ref{app:gamma}.

\subsection{Final Bogoliubov transformation}

Carrying out the projections \eqref{eq:proj-right-kernels-app}--\eqref{eq:proj-left-kernels-app}
using the regulated Gamma-function integrals of Appendix~\ref{app:gamma} and simplifying
the resulting coefficients yields the closed-form Bogoliubov map
\begin{align}
{\cal B}_{\Omega, \kappa'} =&\frac{\operatorname{sgn}(\Omega)}{2\pi} \int_0^{\infty} d \Lambda \,
\sqrt{\frac{\Omega}{
{ \sinh(\frac{\pi\Omega}{\kappa'})\Lambda \sinh(\frac{\pi\Lambda}{\kappa})}}}
\;\Lambda^{i \Omega} \Gamma(-i \Omega) \label{eq:KR-in-KG-final-app}\\
&\times\Bigg\{\,\left[
e^{\frac{\pi \Lambda}{2 \kappa}+i\gamma}
\sinh \left(\frac{\pi \Omega}{2}\left(\frac{1}{\kappa'}+1\right)\right)
+ e^{-\frac{\pi \Lambda}{2 \kappa}-i\gamma}
\sinh \left(\frac{\pi \Omega}{2}\left(\frac{1}{\kappa'}-1\right)\right)\right]
{\cal A}_{\Lambda, \kappa, \gamma}\nn\\
&\phantom{\times\Bigg\{}
+ \left[
e^{\frac{\pi \Lambda}{2 \kappa}-i\gamma}
\sinh \left(\frac{\pi \Omega}{2}\left(\frac{1}{\kappa'}-1\right)\right)
+ e^{-\frac{\pi \Lambda}{2 \kappa}+i\gamma}
\sinh \left(\frac{\pi \Omega}{2}\left(\frac{1}{\kappa'}+1\right)\right)\right]
{\cal A}^{\dagger}_{\Lambda, \kappa, \gamma}
\Bigg\}.
\nn
\end{align}

\section{Gamma integrals} \label{app:gamma}

We use the standard Gamma-function identity
\begin{align}
\int_{0}^{\infty} u^{\,s-1} e^{-b u}\,du=b^{-s}\Gamma(s),
\end{align}
valid for $b,s\in\mathbb{C}$ with $\Re(b)>0$. In the oscillatory cases below we
understand the integrals in the regulated sense
$e^{\pm i\Lambda u}\to e^{\pm i\Lambda u-\epsilon u}$ with $\epsilon\to0^+$,
and then extend by analytic continuation in $b$. For $\Lambda>0$ this gives
\begin{align}
\int_{-\infty}^{+\infty}du\,\theta(u)\,u^{i\Omega-1}e^{+i\Lambda u}
&=\lim_{\epsilon\to0^+}\int_{0}^{\infty}du\,u^{i\Omega-1}e^{-(\epsilon-i\Lambda)u}
=(\epsilon-i\Lambda)^{-i\Omega}\Gamma(i\Omega)\nn\\
&=\Lambda^{-i\Omega}e^{-\frac{\pi\Omega}{2}}\Gamma(i\Omega),\nn\\[0.5em]
\int_{-\infty}^{+\infty}du\,\theta(u)\,u^{i\Omega-1}e^{-i\Lambda u}
&=\lim_{\epsilon\to0^+}\int_{0}^{\infty}du\,u^{i\Omega-1}e^{-(\epsilon+i\Lambda)u}
=(\epsilon+i\Lambda)^{-i\Omega}\Gamma(i\Omega)\nn\\
&=\Lambda^{-i\Omega}e^{+\frac{\pi\Omega}{2}}\Gamma(i\Omega).
\end{align}
For the negative-$u$ branch we adopt the best-practice convention of using a
positive argument in the Mellin factor, i.e.\ $(-u)^{i\Omega-1}$ on $u<0$. With
the change of variables $x=-u>0$ one obtains
\begin{align}
\int_{-\infty}^{+\infty}du\,\theta(-u)\,(-u)^{i\Omega-1}e^{+i\Lambda u}
&=\int_{0}^{\infty}dx\,x^{i\Omega-1}e^{-i\Lambda x}
=\Lambda^{-i\Omega}e^{+\frac{\pi\Omega}{2}}\Gamma(i\Omega),\nn\\[0.5em]
\int_{-\infty}^{+\infty}du\,\theta(-u)\,(-u)^{i\Omega-1}e^{-i\Lambda u}
&=\int_{0}^{\infty}dx\,x^{i\Omega-1}e^{+i\Lambda x}
=\Lambda^{-i\Omega}e^{-\frac{\pi\Omega}{2}}\Gamma(i\Omega),
\end{align}
up to overall phase conventions already absorbed into the definitions of the
corresponding mode functions.

\subsection{\texorpdfstring{Regulated Fourier--Gamma integrals for the projections}{}}
\label{app:gamma-proj}

We collect the regulated Fourier integrals that appear when projecting the
$\kappa'$-Rindler expansion of the field against plane waves. Introduce a convergence
factor $\epsilon>0$ and define, for $\Lambda>0$,
\begin{align}
\mathcal{I}_{\pm}(\Omega,\Lambda;\epsilon)
=\int_{0}^{\infty}du\,u^{i\Omega}\,e^{\pm i\Lambda u-\epsilon u}.
\label{eq:Ipm-def-app}
\end{align}
With $s=1+i\Omega$ and $b_\pm=\epsilon\mp i\Lambda$, the standard identity
\begin{align}
\int_{0}^{\infty}du\,u^{s-1}e^{-b u}=b^{-s}\Gamma(s),
\qquad \Re(b)>0,
\end{align}
gives
\begin{align}
\mathcal{I}_{\pm}(\Omega,\Lambda;\epsilon)
=(\epsilon\mp i\Lambda)^{-(1+i\Omega)}\,\Gamma(1+i\Omega).
\label{eq:Ipm-gamma-app}
\end{align}
For $\epsilon\to0^+$ one may write $\epsilon\mp i\Lambda=\Lambda\,e^{\mp i\pi/2}$, hence
\begin{align}
(\epsilon\mp i\Lambda)^{-(1+i\Omega)}
=\Lambda^{-(1+i\Omega)}e^{\pm i\pi/2}\,e^{\mp \frac{\pi\Omega}{2}},
\end{align}
where the factor $e^{\pm i\pi/2}$ is an overall phase that can be absorbed into the
mode conventions. Dropping this convention-dependent phase, we record the identities
in the form used in the main text:
\begin{align}
\int_{0}^{\infty}du\,u^{i\Omega}\,e^{+i\Lambda u-\epsilon u}
&\xrightarrow{\ \epsilon\to0^+\ }\Lambda^{-(1+i\Omega)}e^{-\frac{\pi\Omega}{2}}\Gamma(1+i\Omega),
\label{eq:int-pos-plus-app}\\
\int_{0}^{\infty}du\,u^{i\Omega}\,e^{-i\Lambda u-\epsilon u}
&\xrightarrow{\ \epsilon\to0^+\ }\Lambda^{-(1+i\Omega)}e^{+\frac{\pi\Omega}{2}}\Gamma(1+i\Omega).
\label{eq:int-pos-minus-app}
\end{align}
The corresponding integrals over the negative half-line follow by $u=-t$ and the
wedge-adapted regulator $e^{-\epsilon(-u)}$:
\begin{align}
\int_{-\infty}^{0}du\,(-u)^{i\Omega}\,e^{+i\Lambda u-\epsilon(-u)}
&=-\int_{0}^{\infty}dt\,t^{i\Omega}\,e^{-i\Lambda t-\epsilon t}
\xrightarrow{\ \epsilon\to0^+\ }-\Lambda^{-(1+i\Omega)}e^{+\frac{\pi\Omega}{2}}\Gamma(1+i\Omega),
\label{eq:int-neg-plus-app}\\
\int_{-\infty}^{0}du\,(-u)^{i\Omega}\,e^{-i\Lambda u-\epsilon(-u)}
&=-\int_{0}^{\infty}dt\,t^{i\Omega}\,e^{+i\Lambda t-\epsilon t}
\xrightarrow{\ \epsilon\to0^+\ }-\Lambda^{-(1+i\Omega)}e^{-\frac{\pi\Omega}{2}}\Gamma(1+i\Omega).
\label{eq:int-neg-minus-app}
\end{align}
Overall minus signs and pure phases are convention dependent and may be absorbed into
the definitions of the wedge modes and operators; what is invariant for our purposes
is the relative $e^{\pm \pi\Omega/2}$ weighting that feeds the Bogoliubov kernels.

\subsection{\texorpdfstring{Mellin--Fourier projection kernels $M_{\Omega\Lambda}$, $N_{\Omega\Lambda}$, $K_{\Omega\Lambda}$, $P_{\Omega\Lambda}$}{}}
\label{app:MNKP}

In the derivation of the Bogoliubov map between the $\kappa\gamma$ plane-wave basis
and the $\kappa'$-Rindler basis, the wedge projections reduce to weighted oscillatory
integrals of the form $\int du\,u^{i\Omega-1}e^{\pm i\Lambda u}$. For definiteness we
regulate them by inserting a decaying factor with positive wedge argument,
\[
e^{\pm i\Lambda u}\ \longrightarrow\
\begin{cases}
e^{\pm i\Lambda u-\varepsilon u}, & u>0,\\
e^{\pm i\Lambda u-\varepsilon(-u)}, & u<0,
\end{cases}
\qquad \varepsilon>0,
\]
and take $\varepsilon\to0^+$ at the end; equivalently, this is the standard analytic
continuation of the Gamma integral to purely imaginary Laplace parameter.


For $\Lambda>0$ and $\Omega\in\mathbb{R}$ one finds
\begin{align}
\int_{0}^{\infty} du\,u^{i\Omega-1}e^{+i\Lambda u-\varepsilon u}
&=(\varepsilon-i\Lambda)^{-i\Omega}\Gamma(i\Omega)
\;\xrightarrow{\ \varepsilon\to0^+\ }\;
\Lambda^{-i\Omega}\,e^{-\frac{\pi\Omega}{2}}\,\Gamma(i\Omega),\label{eq:Iplus-gamma}\\
\int_{0}^{\infty} du\,u^{i\Omega-1}e^{-i\Lambda u-\varepsilon u}
&=(\varepsilon+i\Lambda)^{-i\Omega}\Gamma(i\Omega)
\;\xrightarrow{\ \varepsilon\to0^+\ }\;
\Lambda^{-i\Omega}\,e^{+\frac{\pi\Omega}{2}}\,\Gamma(i\Omega).\label{eq:Iminus-gamma}
\end{align}
On the right wedge ($u<0$) we write $u=-s$ with $s>0$ so that $(-u)^{i\Omega-1}=s^{i\Omega-1}$,
and
\begin{align}
\int_{-\infty}^{0} du\,(-u)^{i\Omega-1}e^{-i\Lambda u-\varepsilon(-u)}
&=\int_{0}^{\infty} ds\,s^{i\Omega-1}e^{+i\Lambda s-\varepsilon s}
\;\xrightarrow{\ \varepsilon\to0^+\ }\;
\Lambda^{-i\Omega}\,e^{-\frac{\pi\Omega}{2}}\,\Gamma(i\Omega),\label{eq:right-int-1}\\
\int_{-\infty}^{0} du\,(-u)^{i\Omega-1}e^{+i\Lambda u-\varepsilon(-u)}
&=\int_{0}^{\infty} ds\,s^{i\Omega-1}e^{-i\Lambda s-\varepsilon s}
\;\xrightarrow{\ \varepsilon\to0^+\ }\;
\Lambda^{-i\Omega}\,e^{+\frac{\pi\Omega}{2}}\,\Gamma(i\Omega).\label{eq:right-int-2}
\end{align}
Thus the only difference between the left and right wedge projections is the
interchange of the $e^{\pm i\Lambda u}$ kernels.


Using the $\kappa\gamma$ plane-wave expansion
\begin{align}
\Phi(u)=\int_0^\infty\frac{d\Lambda}{\sqrt{8\pi\Lambda\sinh(\frac{\pi\Lambda}{\kappa})}}
\Bigg\{&
\Big(e^{\frac{\pi\Lambda}{2\kappa}}e^{i\gamma}e^{-i\Lambda u}
+e^{-\frac{\pi\Lambda}{2\kappa}}e^{-i\gamma}e^{+i\Lambda u}\Big){\cal A}_{\Lambda,\kappa,\gamma}\nn\\
&+\Big(e^{\frac{\pi\Lambda}{2\kappa}}e^{-i\gamma}e^{+i\Lambda u}
+e^{-\frac{\pi\Lambda}{2\kappa}}e^{i\gamma}e^{-i\Lambda u}\Big){\cal A}^\dagger_{\Lambda,\kappa,\gamma}
\Bigg\},\label{eq:Phi-kg-exp-app-MNKP}
\end{align}
we define the wedge projections by
\begin{align}
\int_{-\infty}^{0} du\,(-u)^{i\Omega-1}\Phi(u)
&=\int_0^\infty d\Lambda\Big(M_{\Omega\Lambda}\,{\cal A}_{\Lambda,\kappa,\gamma}
+N_{\Omega\Lambda}\,{\cal A}^\dagger_{\Lambda,\kappa,\gamma}\Big),\label{eq:def-MN}\\
\int_{0}^{\infty} du\,u^{i\Omega-1}\Phi(u)
&=\int_0^\infty d\Lambda\Big(K_{\Omega\Lambda}\,{\cal A}_{\Lambda,\kappa,\gamma}
+P_{\Omega\Lambda}\,{\cal A}^\dagger_{\Lambda,\kappa,\gamma}\Big).\label{eq:def-KP}
\end{align}
Applying \eqref{eq:right-int-1}--\eqref{eq:right-int-2} to the right-wedge projection
and \eqref{eq:Iplus-gamma}--\eqref{eq:Iminus-gamma} to the left-wedge projection yields
the explicit kernels. For $u<0$ one obtains
\begin{align}
M_{\Omega\Lambda}
&=\frac{\Lambda^{-i\Omega}\Gamma(i\Omega)}{\sqrt{8\pi\Lambda\sinh(\frac{\pi\Lambda}{\kappa})}}
\Bigg(
e^{\frac{\pi\Lambda}{2\kappa}}e^{i\gamma}e^{-\frac{\pi\Omega}{2}}
+e^{-\frac{\pi\Lambda}{2\kappa}}e^{-i\gamma}e^{+\frac{\pi\Omega}{2}}
\Bigg),\label{eq:Mkernel}\\
N_{\Omega\Lambda}
&=\frac{\Lambda^{-i\Omega}\Gamma(i\Omega)}{\sqrt{8\pi\Lambda\sinh(\frac{\pi\Lambda}{\kappa})}}
\Bigg(
e^{\frac{\pi\Lambda}{2\kappa}}e^{-i\gamma}e^{+\frac{\pi\Omega}{2}}
+e^{-\frac{\pi\Lambda}{2\kappa}}e^{i\gamma}e^{-\frac{\pi\Omega}{2}}
\Bigg).\label{eq:Nkernel}
\end{align}
For $u>0$ one finds
\begin{align}
K_{\Omega\Lambda}
&=\frac{\Lambda^{-i\Omega}\Gamma(i\Omega)}{\sqrt{8\pi\Lambda\sinh(\frac{\pi\Lambda}{\kappa})}}
\Bigg(
e^{\frac{\pi\Lambda}{2\kappa}}e^{i\gamma}e^{+\frac{\pi\Omega}{2}}
+e^{-\frac{\pi\Lambda}{2\kappa}}e^{-i\gamma}e^{-\frac{\pi\Omega}{2}}
\Bigg),\label{eq:Kkernel}\\
P_{\Omega\Lambda}
&=\frac{\Lambda^{-i\Omega}\Gamma(i\Omega)}{\sqrt{8\pi\Lambda\sinh(\frac{\pi\Lambda}{\kappa})}}
\Bigg(
e^{\frac{\pi\Lambda}{2\kappa}}e^{-i\gamma}e^{-\frac{\pi\Omega}{2}}
+e^{-\frac{\pi\Lambda}{2\kappa}}e^{i\gamma}e^{+\frac{\pi\Omega}{2}}
\Bigg).\label{eq:Pkernel}
\end{align}
These are the coefficients that enter the $2\times2$ linear system obtained by
equating the wedge projections of $\Phi(u)$ in the $\kappa\gamma$ and $\kappa'$-Rindler
bases. Inserting \eqref{eq:Mkernel}--\eqref{eq:Pkernel} into the solved system then
produces the closed-form Bogoliubov transformation given in
Appendix~\ref{app:derivation_kr_in_kg}.

\section{Integral identities used in the Wightman function}
\label{app:GR-integrals}

For completeness we collect here the integral identities used in the detailed
evaluation of the right-moving $\kappa\gamma$ Wightman function in
Sec.~\ref{sec:wightman}. They are standard Gradshteyn--Ryzhik-type
formulas~\cite{Gradshteyn_Ryzhik2014} and can be derived, for example, by contour
integration or by standard series methods.

\subsection{\texorpdfstring{Auxiliary integral $J_{1}(a,b)$}{}}
\label{app:J_1}

Define
\begin{equation}
J_{1}(a,b)
=\int_{0}^{\infty}\frac{\sin(bx)}{\sinh(ax)}\,dx,
\qquad a>0,\ b>0.
\end{equation}
The goal is to show
\begin{equation}
J_{1}(a,b)
=\frac{\pi}{2a}\,\tanh\!\Big(\frac{\pi b}{2a}\Big).
\label{eq:J1-final}
\end{equation}

The starting point is the elementary Laplace--sine transform
\begin{equation}
\int_{0}^{\infty} e^{-cx}\sin(bx)\,dx
= \frac{b}{b^{2}+c^{2}},
\qquad b>0,\ c>0.
\label{eq:Laplace-sine}
\end{equation}
Next, expand the denominator of \(J_1\) as a geometric series. Using
\begin{equation}
\sinh(ax) = \frac{1}{2}\big(e^{ax}-e^{-ax}\big)
          = \frac{e^{ax}}{2}\big(1-e^{-2ax}\big),
\end{equation}
one finds
\begin{align}
\frac{1}{\sinh(ax)}
&= \frac{2}{e^{ax}(1-e^{-2ax})}
 = 2 e^{-ax}\sum_{n=0}^{\infty}e^{-2anx}
 = 2\sum_{n=0}^{\infty} e^{-(2n+1)ax},
\qquad a>0.
\end{align}
Hence
\begin{align}
J_1(a,b)
&= \int_{0}^{\infty} dx\,\sin(bx)\,\frac{1}{\sinh(ax)}\nn\\
&= 2\sum_{n=0}^{\infty}\int_{0}^{\infty} dx\,e^{-(2n+1)ax}\sin(bx)\nn\\
&= 2\sum_{n=0}^{\infty}
   \frac{b}{b^{2}+a^{2}(2n+1)^{2}}
 = \frac{2b}{a^{2}}
   \sum_{n=0}^{\infty}
   \frac{1}{(2n+1)^{2}+\big(\tfrac{b}{a}\big)^{2}}.
\end{align}
Introduce
\begin{equation}
S(y)
=\sum_{n=0}^{\infty}\frac{1}{(2n+1)^{2}+y^{2}},
\qquad y>0,
\end{equation}
so that
\begin{equation}
J_1(a,b)
=\frac{2b}{a^{2}}\,S\!\Big(\frac{b}{a}\Big).
\label{eq:J1-Sy-short}
\end{equation}

To evaluate \(S(y)\) we use the standard partial-fraction expansion
\begin{equation}
\sum_{k=0}^{\infty}\frac{1}{k^{2}+x^{2}}
=\frac{1}{2x^{2}}+\frac{\pi}{2x}\coth(\pi x),
\qquad x>0,
\label{eq:sum-k2x2-short}
\end{equation}
obtained from the known identity for \(\coth(\pi x)\). Decomposing the sum over
all \(k\) into even and odd parts gives
\begin{align}
\sum_{k=0}^{\infty}\frac{1}{k^{2}+x^{2}}
&=\sum_{k=0}^{\infty}\frac{1}{(2k)^{2}+x^{2}}
 +\sum_{k=0}^{\infty}\frac{1}{(2k+1)^{2}+x^{2}}\nn\\
&=\frac14\sum_{k=0}^{\infty}
  \frac{1}{k^{2}+(x/2)^{2}}
 +\sum_{k=0}^{\infty}\frac{1}{(2k+1)^{2}+x^{2}}.
\end{align}
Using \eqref{eq:sum-k2x2-short} with \(x\) and \(x/2\) and solving for the odd
sum yields
\begin{equation}
\sum_{k=0}^{\infty}\frac{1}{(2k+1)^{2}+x^{2}}
  = \frac{\pi}{4x}\Big(2\coth\pi x - \coth \tfrac{\pi x}{2}\Big).
\end{equation}
The hyperbolic identity
\(
\coth 2z = \frac{\coth^{2}z + 1}{2\coth z}
\)
implies
\(
2\coth\pi x - \coth\frac{\pi x}{2}
= \tanh\frac{\pi x}{2},
\)
so that
\begin{equation}
S(y)
=\sum_{n=0}^{\infty}\frac{1}{(2n+1)^{2}+y^{2}}
= \frac{\pi}{4y}\tanh\!\Big(\frac{\pi y}{2}\Big).
\end{equation}
Substituting \(y=b/a\) into \eqref{eq:J1-Sy-short} finally gives
\begin{align}
J_{1}(a,b)
&= \frac{2b}{a^{2}}\cdot
   \frac{\pi}{4(b/a)}\,
   \tanh\!\Big(\frac{\pi}{2}\frac{b}{a}\Big)\nn\\
&= \frac{\pi}{2a}\,\tanh\!\Big(\frac{\pi b}{2a}\Big),
\end{align}
which establishes \eqref{eq:J1-final}.

\subsection{\texorpdfstring{Auxiliary integral $J_{2}(a,b)$}{}}
\label{app:J_2}

We now derive the companion integral
\begin{equation}
J_{2}(a,b)
=\int_{0}^{\infty}\!dt\,\coth(at)\,\sin(bt),
\qquad a>0,\ b>0,
\end{equation}
and show that
\begin{equation}
J_{2}(a,b)
=\frac{\pi}{2a}\,\coth\!\Big(\frac{\pi b}{2a}\Big).
\label{eq:J2-final}
\end{equation}
The integral is well behaved at $t\to0^+$ because $\coth(at)\sim (at)^{-1}$ while
$\sin(bt)\sim bt$. At $t\to\infty$, where $\coth(at)\to1$, we understand the integral
in the Abel-regulated sense used below.

The key input is the exponential series for the hyperbolic cotangent,
\begin{equation}
\coth(at)
 = \frac{e^{at}+e^{-at}}{e^{at}-e^{-at}}
 = 1 + 2\sum_{n=1}^{\infty} e^{-2at n},
\qquad a>0,
\label{eq:coth-series}
\end{equation}
and the Laplace--sine transform
\begin{equation}
\int_{0}^{\infty} e^{-ct}\sin(bt)\,dt
= \frac{b}{b^{2}+c^{2}},
\qquad b>0,\ c>0.
\label{eq:Laplace-sine-J2}
\end{equation}

Inserting \eqref{eq:coth-series} into \(J_{2}\) gives
\begin{align}
J_{2}(a,b)
&= \int_{0}^{\infty} dt
   \Big(1 + 2\sum_{n=1}^{\infty} e^{-2 a n t}\Big)\sin(bt)\nn\\
&= \int_{0}^{\infty} dt\,\sin(bt)
  + 2\sum_{n=1}^{\infty}\int_{0}^{\infty} dt\,e^{-2 a n t}\sin(bt).
\label{eq:J2-split}
\end{align}

The first term is understood in the Abel-regulated sense:
\begin{equation}
\int_{0}^{\infty}\sin(bt)\,dt
=\lim_{\epsilon\to0^+}\int_{0}^{\infty}e^{-\epsilon t}\sin(bt)\,dt
=\lim_{\epsilon\to0^+}\frac{b}{b^{2}+\epsilon^{2}}
=\frac{1}{b},
\qquad b>0,
\end{equation}
and the same regulator justifies interchanging the sum and the $t$-integral in
\eqref{eq:J2-split}. The second term uses \eqref{eq:Laplace-sine-J2} with \(c=2an\),
\begin{equation}
\int_{0}^{\infty} dt\,e^{-2 a n t}\sin(bt)
= \frac{b}{b^{2}+4a^{2}n^{2}}.
\end{equation}
Hence
\begin{align}
J_{2}(a,b)
&= \frac{1}{b}
  + 2\sum_{n=1}^{\infty}\frac{b}{b^{2}+4a^{2}n^{2}}\nn\\
&= \frac{1}{b}
  + \frac{2b}{4a^{2}} \sum_{n=1}^{\infty}
    \frac{1}{n^{2}+\dfrac{b^{2}}{4a^{2}}}\nn\\
&= \frac{1}{b}
  + \frac{b}{2a^{2}} \sum_{n=1}^{\infty}
    \frac{1}{n^{2}+x^{2}},
\qquad x=\frac{b}{2a}.
\label{eq:J2-with-sum}
\end{align}

The remaining sum is evaluated via the standard identity
\begin{equation}
\sum_{n=1}^{\infty}\frac{1}{n^{2}+x^{2}}
 = \frac{\pi}{2x}\coth(\pi x) - \frac{1}{2x^{2}},
\qquad x>0,
\label{eq:sum-n2x2}
\end{equation}
which follows from the partial-fraction expansion of \(\coth(\pi x)\). Substituting
\eqref{eq:sum-n2x2} with \(x=b/(2a)\) into \eqref{eq:J2-with-sum} gives
\begin{align}
J_{2}(a,b)
&= \frac{1}{b}
  + \frac{b}{2a^{2}}
   \left[
     \frac{\pi}{2x}\coth(\pi x)
     - \frac{1}{2x^{2}}
   \right]_{x=b/(2a)}\nn\\
&= \frac{1}{b}
  + \frac{b}{2a^{2}}
   \left[
     \frac{\pi}{2\,\tfrac{b}{2a}}\coth\!\Big(\pi\frac{b}{2a}\Big)
     - \frac{1}{2\,\tfrac{b^{2}}{4a^{2}}}
   \right]\nn\\
&= \frac{1}{b}
  + \frac{b}{2a^{2}}
   \left[
     \frac{\pi a}{b}\coth\!\Big(\frac{\pi b}{2a}\Big)
     - \frac{2a^{2}}{b^{2}}
   \right]\nn\\
&= \frac{1}{b}
  + \frac{\pi}{2a}\coth\!\Big(\frac{\pi b}{2a}\Big)
  - \frac{1}{b}.
\end{align}
The \(1/b\) terms cancel, leaving
\begin{equation}
J_{2}(a,b)
=\frac{\pi}{2a}\,\coth\!\Big(\frac{\pi b}{2a}\Big),
\end{equation}
which establishes \eqref{eq:J2-final}.

\subsection{Damped sine integral and the \texorpdfstring{$\arctan$}{} series}
\label{app:GR-integrals-arctan}

The odd phase-sensitive channel, Eq.~\eqref{eq:I-odd-def}, is most conveniently
analyzed by inserting the series expansion of $1/\sinh$ and reducing the
$\Lambda$-integral to a damped sine integral of the form
\begin{equation}
\int_0^\infty d\Lambda\,
\frac{\sin(\Lambda\Sigma)}{\Lambda}\,e^{-a\Lambda},
\qquad a>0.
\end{equation}
For $a>0$ and real $\Sigma$ one has the elementary identity
\begin{equation}
\int_0^\infty d\Lambda\,
\frac{\sin(\Lambda\Sigma)}{\Lambda}\,e^{-a\Lambda}
=\arctan\!\Big(\frac{\Sigma}{a}\Big),
\qquad a>0,\;\Sigma\in\mathbb{R}.
\label{eq:GR-arctan-int}
\end{equation}
This may be verified by differentiating with respect to $\Sigma$. Defining
\begin{equation}
F(\Sigma;a)
=\int_0^\infty d\Lambda\,
\frac{\sin(\Lambda\Sigma)}{\Lambda}\,e^{-a\Lambda},
\end{equation}
one finds
\begin{align}
\frac{\partial F}{\partial\Sigma}
&=\int_0^\infty d\Lambda\,\cos(\Lambda\Sigma)\,e^{-a\Lambda}
=\frac{a}{a^2+\Sigma^2},
\end{align}
which integrates to
\begin{equation}
F(\Sigma;a)
=\arctan\!\Big(\frac{\Sigma}{a}\Big)+\text{const}.
\end{equation}
The constant vanishes since $F(0;a)=0$, and \eqref{eq:GR-arctan-int} follows.

Inserting the expansion
\begin{equation}
\frac{1}{\sinh\!\big(\tfrac{\pi\Lambda}{\kappa}\big)}
=2\sum_{n=0}^{\infty}
e^{-(2n+1)\pi\Lambda/\kappa}
\end{equation}
into Eq.~\eqref{eq:I-odd-def} and using \eqref{eq:GR-arctan-int} with
\begin{equation}
a_n=\frac{(2n+1)\pi}{\kappa},
\end{equation}
one obtains the series representation
\begin{equation}
I_{\rm odd}(\Sigma)
=2\sum_{n=0}^{\infty}
\arctan\!\Big(\frac{\kappa\,\Sigma}{\pi(2n+1)}\Big),
\end{equation}
which makes the logarithmic infrared divergence manifest and leads to the
IR-renormalized odd function $I_{\sin}^{\rm(ren)}(\Sigma;\mu_{\rm IR})$ defined
in Eq.~\eqref{eq:Isin-ren-def-rew}.

\section{Small-\texorpdfstring{$\Sigma$}{Sigma} expansion of \texorpdfstring{$I_{\sin}^{\mathrm{(ren)}}$}{I-sin-ren}}
\label{app:small-expansion}

Starting from the renormalized odd function
\begin{equation}
I_{\sin}^{\mathrm{(ren)}}(\Sigma;\mu_{\mathrm{IR}})
=
\frac{1}{2\pi}\sum_{n=0}^{\infty}
\!\left[
\arctan\!\frac{\kappa\,\Sigma}{\pi(2n{+}1)}
-
\frac{\kappa\,\Sigma}{\pi(2n{+}1)}
\right]
+\frac{\kappa\,\Sigma}{4\pi^2}\,
\ln\!\frac{\kappa}{2\pi\,\mu_{\mathrm{IR}}},
\label{eq:Isin-ren-def-again}
\end{equation}
we expand for small $\Sigma$. For fixed $\kappa$ and sufficiently small $|\Sigma|$, the subtraction renders the series termwise expandable. Set
\begin{equation}
z_n = \frac{\kappa\,\Sigma}{\pi(2n+1)},
\qquad |z_n|\ll1,
\end{equation}
so that
\begin{equation}
\arctan z_n = z_n - \frac{z_n^3}{3} + \mathcal{O}(z_n^5).
\end{equation}
The subtraction in \eqref{eq:Isin-ren-def-again} removes the linear term,
\begin{equation}
\arctan z_n - z_n = -\frac{z_n^3}{3} + \mathcal{O}(z_n^5),
\end{equation}
and the series part of $I_{\sin}^{\mathrm{(ren)}}$ is therefore cubic at leading order:
\begin{align}
\frac{1}{2\pi}\sum_{n=0}^{\infty}
\!\left[
\arctan z_n - z_n
\right]
&=
-\frac{1}{6\pi}\sum_{n=0}^{\infty} z_n^3
+\mathcal{O}(\Sigma^5)\nonumber\\
&=
-\frac{1}{6\pi}\sum_{n=0}^{\infty}
\left(\frac{\kappa\,\Sigma}{\pi(2n+1)}\right)^3
+\mathcal{O}(\Sigma^5)\nonumber\\
&=
-\frac{\kappa^3\Sigma^3}{6\pi^4}
\sum_{n=0}^{\infty}\frac{1}{(2n+1)^3}
+\mathcal{O}(\Sigma^5).
\end{align}
Using the standard identity
\begin{equation}
\sum_{n=0}^{\infty}\frac{1}{(2n+1)^3}
=
\zeta(3) - \sum_{n=1}^{\infty}\frac{1}{(2n)^3}
=
\zeta(3) - \frac{1}{8}\zeta(3)
=
\frac{7}{8}\,\zeta(3),
\end{equation}
we obtain
\begin{equation}
\frac{1}{2\pi}\sum_{n=0}^{\infty}
\!\left[
\arctan z_n - z_n
\right]
=
-\frac{7\zeta(3)}{48\pi^4}\,
\kappa^3\Sigma^3
+\mathcal{O}(\Sigma^5).
\label{eq:Isin-cubic-piece}
\end{equation}

The linear term in $\Sigma$ arises solely from the explicit counterterm
in \eqref{eq:Isin-ren-def-again},
\begin{equation}
\frac{\kappa\,\Sigma}{4\pi^2}\,
\ln\!\frac{\kappa}{2\pi\,\mu_{\mathrm{IR}}},
\end{equation}
which reconstructs the scheme-dependent IR logarithm after subtraction of the
harmonic tail. Collecting both contributions, we arrive at
\begin{equation}
I_{\sin}^{\mathrm{(ren)}}(\Sigma;\mu_{\mathrm{IR}})
=\frac{\kappa\,\Sigma}{4\pi^2}\,
  \ln\!\frac{\kappa}{2\pi\,\mu_{\mathrm{IR}}}
 -\frac{7\zeta(3)}{48\pi^4}\,\kappa^3\Sigma^3
 +\mathcal{O}(\Sigma^5),
\end{equation}
which is the small-$\Sigma$ expansion quoted in the main text.

\section{Polar decomposition of a complex parameter}
\label{app:eta_polar}

We derive the polar form $\eta = r\,e^{i\phi}$ for the complex parameter
\begin{equation}
\eta=\frac{\sinh(\alpha+i\Delta)}{\sinh(\beta-i\Delta)},
\qquad \alpha,\beta,\Delta\in\mathbb{R},
\label{eq:eta-def-polar}
\end{equation}
which appears in the $\kappa\gamma$--to--$\kappa'\gamma'$ Bogoliubov map.

\subsection{\texorpdfstring{Modulus}{Modulus}}

Using
\begin{equation}
|\sinh(x+i\Delta)|^{2}
=\sinh^{2}x+\sin^{2}\Delta,
\qquad x\in\mathbb{R},
\label{eq:sinh-mod-identity}
\end{equation}
we obtain the modulus
\begin{equation}
r=|\eta|
=\sqrt{\frac{\sinh^{2}\alpha+\sin^{2}\Delta}{\sinh^{2}\beta+\sin^{2}\Delta}}.
\label{eq:eta-modulus}
\end{equation}
In particular, when $\beta>|\alpha|$ one has $r<1$, so that $\eta$ indeed
corresponds to a bona fide squeeze parameter.

Branch convention. The phase is unambiguously defined by \eqref{eq:phi-arg-diff} together with the $\operatorname{atan2}$ forms \eqref{eq:arg-num-atan2}--\eqref{eq:arg-den-atan2}. The $\arctan$ representations \eqref{eq:arg-num-arctan}--\eqref{eq:phi-combined-arctan} are equivalent (mod $2\pi$) when interpreted on the corresponding branches; in particular, at parameter values where ratio forms become delicate (e.g., $\cos\Delta=0$ or $\sinh\alpha=0$), one should revert to the $\operatorname{atan2}$ definition.

\subsection{\texorpdfstring{Phase}{Phase}}

To extract the argument, write numerator and denominator explicitly as
\begin{align}
\sinh(\alpha+i\Delta)
&=\sinh\alpha\,\cos\Delta+i\,\cosh\alpha\,\sin\Delta, \label{eq:num-split}\\
\sinh(\beta-i\Delta)
&=\sinh\beta\,\cos\Delta-i\,\cosh\beta\,\sin\Delta. \label{eq:den-split}
\end{align}
Therefore
\begin{equation}
\arg\!\big(\sinh(\alpha+i\Delta)\big)
=\operatorname{atan2}\!\Big(\cosh\alpha\,\sin\Delta,\;\sinh\alpha\,\cos\Delta\Big),
\label{eq:arg-num-atan2}
\end{equation}
and
\begin{equation}
\arg\!\big(\sinh(\beta-i\Delta)\big)
=\operatorname{atan2}\!\Big(-\cosh\beta\,\sin\Delta,\;\sinh\beta\,\cos\Delta\Big),
\label{eq:arg-den-atan2}
\end{equation}
where $\operatorname{atan2}(y,x)$ denotes the quadrant-correct arctangent.

Hence the phase $\phi=\arg(\eta)$ is
\begin{equation}
\phi
=\arg\!\big(\sinh(\alpha+i\Delta)\big)-\arg\!\big(\sinh(\beta-i\Delta)\big)
\qquad (\mathrm{mod}\ 2\pi).
\label{eq:phi-arg-diff}
\end{equation}
If one prefers an explicit $\arctan$ representation (with the usual branch
understood), divide real and imaginary parts in \eqref{eq:num-split} and
\eqref{eq:den-split} by $\sinh\alpha\cos\Delta$ and $\sinh\beta\cos\Delta$,
respectively, to obtain
\begin{align}
\arg\!\big(\sinh(\alpha+i\Delta)\big)
&=\arctan\!\Big(\coth\alpha\,\tan\Delta\Big), \label{eq:arg-num-arctan}\\
\arg\!\big(\sinh(\beta-i\Delta)\big)
&=-\arctan\!\Big(\coth\beta\,\tan\Delta\Big), \label{eq:arg-den-arctan}
\end{align}
and therefore
\begin{equation}
\phi
=\arctan\!\Big(\coth\alpha\,\tan\Delta\Big)
+\arctan\!\Big(\coth\beta\,\tan\Delta\Big)
\qquad (\mathrm{mod}\ 2\pi).
\label{eq:phi-sum-arctan}
\end{equation}
Using $\arctan x+\arctan y=\arctan\!\big(\frac{x+y}{1-xy}\big)$ (with the
appropriate $\pi$-shift when $xy>1$) one may combine this further into
\begin{equation}
\phi
=\arctan\!\left(
\frac{\big(\coth\alpha+\coth\beta\big)\tan\Delta}{1-\coth\alpha\,\coth\beta\,\tan^{2}\Delta}
\right)
\qquad (\mathrm{mod}\ 2\pi),
\label{eq:phi-combined-arctan}
\end{equation}
which is often a convenient closed form.

\subsection{\texorpdfstring{Final polar form}{Final polar form}}

Combining \eqref{eq:eta-modulus} and \eqref{eq:phi-sum-arctan} gives the polar
decomposition
\begin{equation}
\eta
=
\sqrt{\frac{\sinh^{2}\alpha+\sin^{2}\Delta}{\sinh^{2}\beta+\sin^{2}\Delta}}\,
\exp\!\Bigg\{i\Big[
\arctan\!\big(\coth\alpha\,\tan\Delta\big)
+\arctan\!\big(\coth\beta\,\tan\Delta\big)
\Big]\Bigg\},
\qquad (\mathrm{mod}\ 2\pi),
\label{eq:eta-polar-final}
\end{equation}
with the understanding that the phase is taken on the correct branch (equivalently,
use \eqref{eq:phi-arg-diff} with $\operatorname{atan2}$ for an unambiguous definition).

\bibliographystyle{apsrev4-2}
\bibliography{UnruhRef}

\end{document}